\documentclass[twocolumn]{aastex63}
\usepackage[caption=false]{subfig}
\usepackage{dcolumn}

\shorttitle{FRAMEx I: Volume-Complete Radio and X-ray Snapshot}
\shortauthors{Fischer et al.}

\begin{document}

\title{Fundamental Reference AGN Monitoring Experiment (FRAMEx) I:\\[0.05cm] 
Jumping Out of the Plane with the VLBA}

\correspondingauthor{Travis Fischer}
\email{tfischer@stsci.edu}

\author[0000-0002-3365-8875]{Travis C. Fischer}
\affiliation{AURA for ESA, Space Telescope Science Institute, Baltimore, MD, USA, 3700 San Martin Drive, Baltimore, MD 21218, USA}
\affiliation{U.S. Naval Observatory, 3450 Massachusetts Ave NW, Washington, DC 20392-5420, USA}

\author[0000-0002-4902-8077]{Nathan J. Secrest}
\affiliation{U.S. Naval Observatory, 3450 Massachusetts Ave NW, Washington, DC 20392-5420, USA}

\author[0000-0002-4146-1618]{Megan C. Johnson}
\affiliation{U.S. Naval Observatory, 3450 Massachusetts Ave NW, Washington, DC 20392-5420, USA}

\author[0000-0002-5604-5254]{Bryan N. Dorland}
\affiliation{U.S. Naval Observatory, 3450 Massachusetts Ave NW, Washington, DC 20392-5420, USA}

\author[0000-0002-8736-2463]{Phillip J. Cigan}
\affiliation{U.S. Naval Observatory, 3450 Massachusetts Ave NW, Washington, DC 20392-5420, USA}

\author[0000-0002-0819-3033]{Luis C. Fernandez}
\affiliation{Department of Physics and Astronomy, George Mason University, MS3F3, 4400 University Drive, Fairfax, VA 22030, USA}

\author[0000-0001-8587-9285]{Lucas R. Hunt}
\affiliation{U.S. Naval Observatory, 3450 Massachusetts Ave NW, Washington, DC 20392-5420, USA}

\author[0000-0003-4327-0548]{Michael Koss}
\affiliation{Eureka Scientific, 2452 Delmer Street Suite 100, Oakland, CA 94602-3017, USA}

\author[0000-0001-7376-8481]{Henrique R. Schmitt}
\affiliation{Naval Research Laboratory, Washington, DC 20375, USA}

\author[0000-0002-4873-0972]{Norbert Zacharias}
\affiliation{U.S. Naval Observatory, 3450 Massachusetts Ave NW, Washington, DC 20392-5420, USA}

\begin{abstract}

We present the first results from the Fundamental Reference AGN Monitoring Experiment (FRAMEx), an observational campaign dedicated to understanding the physical processes that affect the apparent positions and morphologies of AGNs. In this work, we obtained simultaneous Swift X-ray Telescope (XRT) and Very Long Baseline Array (VLBA) radio observations for a snapshot campaign of 25 local AGNs that form a volume-complete sample with hard X-ray (14--195~keV) luminosities above $10^{42}$~erg~s$^{-1}$, out to a distance of 40~Mpc. Despite achieving an observation depth of $\sim20$~$\mu$Jy, we find that 16 of 25 AGNs in our sample are not detected with the VLBA on milli-arcsecond (sub-parsec) scales, and the corresponding core radio luminosity upper limits are systematically below predictions from the Fundamental Plane of black hole activity. Using archival Jansky Very Large Array (VLA) radio measurements, our sample jumps back onto the Fundamental Plane, suggesting that extended radio emission is responsible for the apparent correlation between radio emission, X-ray emission, and black hole mass. We suggest that this discrepancy is likely due to extra-nuclear radio emission produced via interactions between the AGN and host environment. We compare VLBA observations of AGNs to VLA observations of nearby Galactic black holes (GBHs) and we find a mass-independent correlation between radio and X-ray luminosities of black holes of $L_\mathrm{6~cm}$/$L_\mathrm{2-10~keV}$ $\sim$ 10$^{-6}$, in line with predictions for coronal emission, but allowing for the possibility of truly radio silent AGNs.

\end{abstract}


\section{Introduction} \label{section: Introduction}
Over the past several decades, supermassive black holes (SMBHs), which manifest as active galactic nuclei (AGNs), have gained profound importance to several fields of astrophysics. With the discovery of the $M_\mathrm{BH}-\sigma$ relation \citep{2000ApJ...539L...9F, 2000ApJ...539L..13G} at the turn of the millennium, it became clear that SMBHs and their host galaxies co-evolve in some fashion, likely involving the mutual feeding of star formation and SMBH accretion, combined with ``feedback'' from the AGN into the surrounding interstellar medium via either the deposition of mechanical energy from a collimated jet, or radiative energy in the form of AGN winds \citep[for an essential review of this topic, see][]{2013ARA&A..51..511K}. Similarly, the discovery of increasingly high-redshift quasars, such as the recent announcement of one powered by an $8\times10^8$~$M_\sun$ SMBH at a redshift of 7.54 \citep{2018Natur.553..473B} when the universe was only a few hundred million years old, has raised serious questions about the formation mechanisms of black holes in the early universe \citep[e.g.,][]{2010A&ARv..18..279V}, and the distribution of quasars across the sky may even provide independent tests of our cosmological assumptions \citep[e.g.,][and references therein]{2019MNRAS.488L.104S}. 

The nearly uniform distribution of quasars across the sky, combined with their negligible proper motions and significant luminosity at almost every wavelength, have additionally made them ideal objects with which to realize a quasi-inertial celestial reference frame, with the first quasar-based realization of the International Celestial Reference System (ICRS), the International Celestial Reference Frame \citep[ICRF;][]{1998AJ....116..516M}, being adopted at the same time as the discovery of SMBH-galaxy scaling relations. Successive refinements of the ICRF have improved both the density of these reference sources and their respective astrometric precision \citep[ICRF2:][and ICRF3: Charlot et al., in prep.]{2015AJ....150...58F}.  The ICRF has historically been a radio reference frame with quasar positions defined at 8.6 GHz.  However, for the first time, the ICRF3 is now multi-wavelength with positions defined at 24 GHz (1.2 cm) and 32 GHz (0.9 cm) \citep[Charlot et al., in prep.][]{}.  
If the ICRF is to be viewed as \emph{the} fundamental reference frame for all astronomical positions and proper motions, there should be no significant disagreement between the radio positions of ICRF sources and their counterparts at other wavelengths. This has unfortunately not been the case, with evidence of the existence of significant optical-radio positional offsets being published soon after the release of ICRF1 \citep{2002AJ....124..612D}, ICRF2  \citep{2013MNRAS.430.2797A,2013A&A...553A..13O,2014AJ....147...95Z} and, along with the second data release of Gaia, ICRF3 \citep{2017ApJ...835L..30M}. 

Earlier, pre-Gaia work hinted at an astrophysical nature to optical-radio offsets \citep{2014AJ....147...95Z}, but the unprecedented astrometric precision of Gaia has put the astrophysical nature of these offsets on firm empirical ground \citep[e.g.,][]{2016AJ....152..118H,2017A&A...598L...1K,2019MNRAS.482.3023P,2019ApJ...871..143P}. It is therefore imperative to understand how the physical phenomena associated with AGN activity, such as jets, intrinsic variability, variable obscuration, binary SMBHs, host galaxy morphology, etc.\ affect the apparent positions of AGNs and quasars across the electromagnetic spectrum. 

Towards this goal, the U.S.\ Naval Observatory and its collaborators from partner institutions have begun the Fundamental Reference AGN Monitoring Experiment (FRAMEx), a research framework that includes the physical characterization and monitoring of AGN phenomena that may affect the apparent positions of AGNs at different wavelengths (see \citet{2020arXiv200902169D} for a more detailed overview of the FRAMEx collaboration and discussion of the relevant angular scales). While direct studies of the quasars that comprise the ICRF are an essential part of FRAMEx, their typical redshifts of 1--2 correspond to a minimum distance separation of $\sim8$ parsec at the milli-arcsecond (mas) resolution of very long baseline interferometry (VLBI) networks such as the Very Long Baseline Array (VLBA). The accretion regions of AGNs, which are typically confined to within $\sim1$~parsec, are therefore much smaller in angular extent than what is typically observable, limiting research to AGN emission on large scales. It is therefore worthwhile to study more nearby AGNs, where mas-scale observations taken with facilities such as the VLBA can resolve features subtending only a few tenths of a parsec, providing a direct probe into the accretion physics of AGNs.

In this work, we present results from an initial snapshot survey of a volume-complete sample of 25 nearby ($<40$~Mpc) AGNs, in which we obtained simultaneous observations with the VLBA and the Neil Gehrels Swift Observatory X-ray Telescope (XRT), as part of a long-term program to monitor the behavior of AGNs at sub-parsec scales. The simultaneity of these data mitigates the uncertainty introduced by variability that is inherent to studying AGNs on small physical scales, allowing a more robust assessment of the relationship between emission mechanisms at different wavelengths and energies.  Our sample was selected from the Swift 105-month Burst Alert Telescope (BAT) catalog \citep{2018ApJS..235....4O}, enabling obscuration-independent selection of a statistically representative sample of nearby AGNs above some intrinsic luminosity limit.

The primary focus of this work is the relationship between the radio luminosity, X-ray luminosity, and black hole mass of AGNs, i.e.\ the Fundamental Plane of black hole activity \citep[e.g.,][hereafter the FP]{2003MNRAS.345.1057M}, which potentially serves as a parameter space unifying black hole accretion across all black hole masses, including Galactic black holes. We additionally explore the relationship between the core radio and X-ray luminosities of these AGNs more broadly, considering the potential bimodality between radio-loud and radio-quiet AGNs, as well as the prevalence of potentially truly radio-silent AGNs. 

\section{Methodology} \label{section: Methodology}
\subsection{Sample Selection} \label{subsection: Sample Selection}
Our AGN sample was selected from the \emph{Swift} BAT 105-month catalog \citep{2018ApJS..235....4O}. Because this catalog has a uniform flux limit ($\sim8\times10^{-12}$~erg~cm$^{-2}$~s$^{-1}$) across the sky, and hard X-rays (14--195~keV) reliably pick out bona-fide AGNs, even in the presence of heavy absorption ($N_H\lesssim10^{24}$~cm$^{-2}$), constructing a volume-complete sample of AGNs above some luminosity threshold is straightforward. Additionally, AGNs from the BAT catalog have extensive multi-wavelength coverage, and have the benefit of intense recent study \citep[e.g.,][]{2017MNRAS.470..800T, 2018ApJ...856..154S, 2018ApJ...858..110P, 2019MNRAS.489.3073B}.

We choose AGNs above a hard X-ray luminosity of $10^{42}$~erg~s$^{-1}$, which corresponds to a distance limit of $\sim40$~Mpc for a volume-complete sample using a flat $\Lambda$CDM cosmology with $H_0=70$~km~s$^{-1}$~Mpc$^{-1}$ and $\Omega_\mathrm{M}=0.3$. This luminosity threshold was chosen to sample both moderate and high luminosity AGNs, and to provide a sample of AGNs close enough to give exquisite physical resolution on the AGN. Additionally, heavy, Compton-thick obscuration may cause lower luminosity AGNs beyond $\sim40$~Mpc to be undetected even by BAT \citep[e.g.,][]{2019ApJ...887..173L}, biasing a sample extending further than $40$~Mpc towards less obscured AGNs. At distances of less than 40 Mpc, the $\sim$1~mas beam of the VLBA at 6~cm (C-band), samples physical scales of less than 0.2~pc, within the dusty obscuring medium (i.e., the ``torus'') inferred to exist in most AGNs \citep[e.g.,][]{2015ARA&A..53..365N}. Additionally, these physical scales correspond to light crossing times of only a few months, enabling relatively short-term studies of both luminosity and morphological variability.

\begin{figure}
\includegraphics[width=\columnwidth]{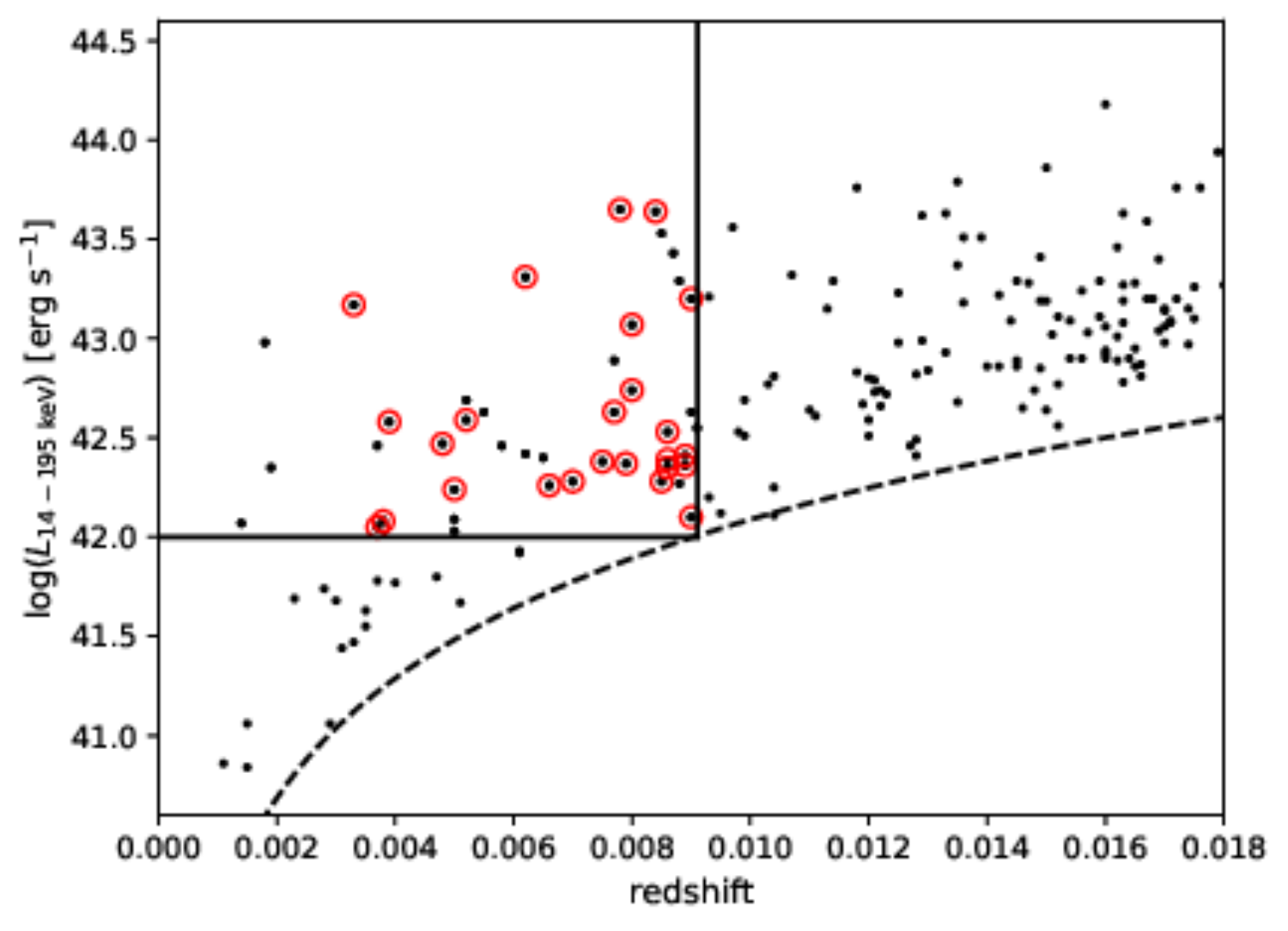}
\caption{Redshift versus hard X-ray luminosity for AGNs from the 105-month Swift BAT catalog, showing the flux limit (dashed line) and our volume-complete cut (solid lines). Sources with declination between $-30\degr$ and $+60\degr$, circled in red, are observable with northern hemisphere facilities such as the VLBA and are the sample explored in this study.}
\label{fig: malmquist}
\end{figure}

Out of the 1105 AGNs in the 105-month BAT catalog, 43 are within the local volume defined here. For observability with the VLBA and other northern hemisphere facilities, we made an additional declination cut of $-30\degr<\delta<+60\degr$, leaving 25 targets. We illustrate our selection method in Figure~\ref{fig: malmquist} and provide a list of targets in Table~\ref{tab:sample}.

Table~\ref{tab:sample} also presents reddening-corrected nuclear narrow H$\alpha$ luminosities, maximum expected star formation rates ($\mathrm{SFR_{max}}$) and maximum expected supernova rate ($\mathrm{SNR_{max}}$). The H$\alpha$ luminosities were calculated using nuclear emission line fluxes from \cite{2017ApJ...850...74K}, except for NGC~1320 \citep{1992A&AS...96..389D}, NGC~2782 \citep{1997ApJS..112..315H} and NGC~7465 \citep{2006ApJS..164...81M}. These flux measurements were obtained with small apertures (2\arcsec\ to 4\arcsec) and corrected for reddening using the \cite{2000ApJ...533..682C} extinction law. These H$\alpha$ luminosities are used to determine the maximum expected supernova rate due to star formation, assuming that all of the narrow H$\alpha$ is due to star formation, providing an upper limit on any possible contamination of the nuclear AGN radio emission by SNe. We calculated $\mathrm{SFR_{max}}$ for our sample using the calibration from \cite{1998ARA&A..36..189K}. These values were converted to $\mathrm{SNR_{max}}$ using \textsc{starburst99} continuous star formation models \citep{1999ApJS..123....3L}, assuming a \cite{1955ApJ...121..161S} initial mass function and an age higher than 30~Myr where the supernova rate reaches a plateau. We find $\mathrm{SNR_{max}}$ ranging from 3.5$\times10^{-4}$ to 5.9$\times10^{-2}$ yr$^{-1}$, which allows us to conclude that the contribution from SNe to the nuclear radio emission is negligible.

\begin{deluxetable*}{lrrlccclll} \label{tab:volume}
\tablehead{\colhead{Target} & \colhead{R.A.\ (ICRS)} & \colhead{Decl.\ (ICRS)}  & \colhead{Type} & \colhead{Redshift} & \colhead{Distance} & \colhead{log($M_\mathrm{BH}$)} & \colhead{log($L_\mathrm{H\alpha}$)} & \colhead{$\mathrm{SFR_{max}}$} & \colhead{$\mathrm{SNR_{max}}$}\\ 
[-0.3cm]
& \colhead{(deg)}  & \colhead{(deg)}    &  & & \colhead{(Mpc)} & \colhead{[$M_{\sun}$]} & \colhead{[erg~s$^{-1}$]} & \colhead{(M$_{\odot}$~yr$^{-1}$)} & \colhead{(yr$^{-1}$)}}
\startdata
NGC 1052 & 40.2699884 & $-$8.25576190   & Sy2   & 0.0050 & 21.5 & 8.67 & 40.15 & 0.11 & 2.2$\times10^{-3}$\\
NGC 1068 & 40.6696342 & $-$0.01323785   & Sy2   & 0.0038 & 16.3 & 6.93 & 41.34 & 1.74 & 3.5$\times10^{-2}$\\
NGC 1320 & 51.2028681 & $-$3.04226840   & Sy2   & 0.0089 & 38.4 & 7.96 & 40.49 & 0.24 & $ 4.9\times10^{-3}$\\
NGC 2110 & 88.0473918 & $-$7.45625094   & Sy2   & 0.0078 & 33.6 & 9.38 & 40.68 & 0.38 & $ 7.5\times10^{-3}$\\
NGC 2782 & 138.5212787 & $+$40.11369022   & Sy2   & 0.0085 & 36.6 & 6.07 & 41.51 & 2.62 & $ 5.2\times10^{-2}$\\
IC 2461  & 139.9914308 & $+$37.19100007   & Sy2   & 0.0075 & 32.3 & 7.27 & 39.40 & 0.02 & $ 3.9\times10^{-4}$\\
NGC 2992 & 146.4247756 & $-$14.32626689   & Sy1   & 0.0077 & 33.2 & 8.33 & 40.11 & 0.10 & $ 2.0\times10^{-3}$\\
NGC 3081 & 149.8731005 & $-$22.82631476   & Sy2   & 0.0080 & 34.5 & 7.74 & 40.97 & 0.74 & $ 1.5\times10^{-2}$\\
NGC 3089 & 149.9028701 & $-$28.33129443   & Sy2?  & 0.0090 & 38.8 & 6.55 & \nodata & \nodata &  \nodata \\
NGC 3079 & 150.4908469 & $+$55.67979744   & Sy2   & 0.0037 & 15.9 & 6.38 & 40.04 & 0.09 & $ 1.7\times10^{-3}$\\
NGC 3227 & 155.8774015 & $+$19.86505766   & Sy1   & 0.0039 & 16.8 & 6.77 & 40.78 & 0.48 & $ 9.6\times10^{-3}$\\
NGC 3786 & 174.9271391 & $+$31.90942732   & Sy2   & 0.0089 & 38.4 & 7.48 & 40.73 & 0.42 & $ 8.4\times10^{-3}$\\
NGC 4151 & 182.6357547 & $+$39.40584860   & Sy1   & 0.0033 & 14.2 & 7.55 & 41.21 & 1.27 & $ 2.5\times10^{-2}$\\
NGC 4180 & 183.2626924 & $+$7.03891255   & LINER & 0.0070 & 30.1 & 7.63 & \nodata & \nodata &  \nodata\\
NGC 4235 & 184.2911678 & $+$7.19157597   & Sy1   & 0.0080 & 34.5 & 7.55 & 40.94 & 0.68 & 1.4$\times10^{-2}$\\
NGC 4388 & 186.4449188 & $+$12.66215153   & Sy2   & 0.0084 & 36.2 & 6.94 & 41.57 & 2.93 & 5.9$\times10^{-2}$\\
NGC 4593 & 189.9143400 & $-$5.34417010   & Sy1   & 0.0090 & 38.8 & 6.88 & 40.65 & 0.35 & 7.0$\times10^{-3}$\\
NGC 5290 & 206.3297085 & $+$41.71241871   & Sy2   & 0.0086 & 37.1 & 7.78 & 40.12 & 0.10 & 2.1$\times10^{-3}$\\
NGC 5506 & 213.3119888 & $-$3.20768334   & Sy1.9 & 0.0062 & 26.7 & 6.96 & 41.18 & 1.19 & 2.4$\times10^{-2}$\\
NGC 5899 & 228.7634964 & $+$42.04991289   & Sy2   & 0.0086 & 37.1 & 7.69 & 41.19 & 1.23 & 2.5$\times10^{-2}$\\
NGC 6814 & 295.6690092 & $-$10.32345792   & Sy1   & 0.0052 & 22.4 & 7.04 & 39.35 & 0.02 & 3.5$\times10^{-4}$\\
NGC 7314 & 338.9424567 & $-$26.05043820   & Sy1.9 & 0.0048 & 20.6 & 6.76 & 39.68 & 0.04 & 7.5$\times10^{-4}$\\
NGC 7378 & 341.9486864 & $-$11.81658744   & Sy2   & 0.0086 & 37.1 & 4.93 & \nodata & \nodata & \nodata\\
NGC 7465 & 345.5039963 & $+$15.96477472   & Sy2   & 0.0066 & 28.4 & 6.54 & 40.44 & 0.22 & 4.3$\times10^{-3}$\\
NGC 7479 & 346.2359605 & $+$12.32295297   & Sy2   & 0.0079 & 34.0 & 7.61 & 40.02 & 0.08 & 1.7$\times10^{-3}$\\
\enddata
\caption{FRAMEx Volume-complete Sample}
\label{tab:sample}
\end{deluxetable*}

We note that the distances we use are calculated from the sample redshifts, and the mean redshift of the sample is 0.007, so the effect of peculiar motions on the cosmological distance estimates may be significant. To estimate this, we retrieved redshift-independent distances for our sample using the NASA Extragalactic Database (NED),\footnote{\url{https://ned.ipac.caltech.edu}} and calculated the amount of additional, intrinsic uncertainty that is required for the luminosity using the cosmological distance estimate to be consistent with the luminosity using the redshift-independent distance, in a reduced $\chi^2$ sense. We found 13 unique methods used to estimate the redshift-independent distances of our sample: ring diameter \citep{1981ApJS...45..541P}, Faber-Jackson and ``tertiary'' methods \citep[e.g.,][]{1984ApJS...56...91D}, Tully estimate distances \citep{1988cng..book.....T}, CO ring diameter \citep{1991PASJ...43..671S}, D-sigma and the Infrared Astronomical Satellite standard candle \citep{1997ApJS..109..333W}, the ``fundamental plane'' of early-type galaxies \citep[unrelated to the fundamental plane of black hole activity; e.g.,][]{2001MNRAS.327.1004B} look-alike (``sosie'') galaxies \citep{2002A&A...393...57T}, surface brightness fluctuations \citep[e.g.,][]{2013AJ....146...86T}, AGN time lags \citep[e.g.,][]{2014Natur.515..528H}, Tully-Fisher \citep[e.g.,][]{2016AJ....152...50T}, and type Ia supernovae \citep[e.g.,][]{2017ApJ...842L..13K}. We found the value the intrinsic uncertainty to be $0.31$~dex. Using the \citet{2003MNRAS.345.1057M} FP relation and propagation of uncertainty, however, the scatter induced by the absolute luminosity uncertainty is only 0.19~dex. Moreover, this source of uncertainty also affected the sources studied by \citet[][see their Section 3.1]{2003MNRAS.345.1057M}, and makes only a minor contribution to the overall dispersion in the FP \citep[see Section 6.1 in][]{2003MNRAS.345.1057M}. We therefore use the cosmological distances for our sample, which are more generally available than redshift-independent distances.

\subsection{VLBA Observations}  \label{subsection: VLBA}
In order to estimate the required integration time of our sample, we used the FP given in \citep{2003MNRAS.345.1057M}, a 10$^6$\,M$_{\odot}$ black hole with a hard X-ray luminosity of 10$^{42}$\,erg\,s$^{-1}$ would result in a 6~cm (6~GHz) radio peak luminosity of approximately 10$^{36}$\,erg\,s$^{-1}$. This yields a flux of $\sim$100\,$\mu$Jy beam$^{-1}$ at the maximum distance of $\sim$40 Mpc for our sample.  To ensure our targets are robustly detected, we designed our observations to achieve a signal-to-noise (S/N) ratio of 5, or a root mean square (RMS) noise of $\sim$20~$\mu$Jy beam$^{-1}$. Using the European VLBI Network's online calculator,\footnote{\url{http://old.evlbi.org/cgi-bin/EVNcalc}} we determined that one hour of on-source integration time produces a $1\sigma$ image thermal noise of $23\mu$Jy~beam$^{-1}$.


To ensure accurate VLBA pointing for all of our targets, which is necessary due to the small effective field of view (a few arcsec) in the C-band, we used the most accurate AGN positions of our targets. For 17 of our objects we determined the AGN positions from archival VLA A-array configuration observations in the X-band ($\sim$9 GHz), which provided a pointing accuracy of roughly 50 mas.  Two of our targets had archival C-band ($\sim$5 GHz) VLA data in the A-array and B-array configurations for an a priori positional accuracy of 80 mas and 370 mas, respectively. The remaining 6 objects did not have any archival VLA data with the required positional accuracy for VLBA pointing and thus, we used the Pan-STARRS $y$-band data for 5 galaxies providing a pointing accuracy of $\sim$ 300 mas and for 1 galaxy we used the Chandra X-ray position, which has an a priori positional accuracy of 500 mas. These AGN positions are well within the desired pointing accuracy for the VLBA and provide high confidence for ensuring we imaged the AGN at our sub-parsec spatial scale should it exist.

 
\begin{deluxetable*}{lccrrclr} \label{tab:radio_obs}
\tablehead{\colhead{Target} & \colhead{Frequency} & \colhead{Restoring Beam}                 & \colhead{Beam angle} & \colhead{RMS} & \colhead{Calibrator} & \colhead{R.A.} & \colhead{Decl.}\\  [-0.2cm]
& \colhead{(GHz)}       &  \colhead{($\alpha \times \delta$; mas)}	& \colhead{(deg)} & \colhead{($\mu$Jy bm$^{-1}$)}  & \colhead{IERS Name} & (deg($\mu$s)) & (deg($\mu$as))}
\startdata
NGC 1052 & 5.873692 & 3.83$\times$1.49 &   8.4 & 4242.0 & 0240$-$060 & 40.801956142(4.78) & $-$5.8486932(137.3) \\
NGC 1068 & 5.858862 & 3.48$\times$1.60 &  14.7 &  21.7 & 0237$-$027  & 39.939467802(2.09)	 & $-$2.57803183(31.9)\\
NGC 1320 & 5.867662 & 7.46$\times$2.02 &$-$15.9 &  39.1 & 0319$-$056  & 50.499459824(8.17) & $-$5.4367857(222.2)\\
NGC 2110 & 5.861240 & 3.37$\times$1.51 &   5.3 &  162.0 & 0551$-$086  & 88.42454680(13.99)	 & $-$8.6671937(157.8) \\
NGC 2782 & 5.867662 & 2.82$\times$2.23 &  18.1 &  33.0 & 0913+391  & 139.203769060(7.04) &  38.9078184(115.6) \\
IC 2461  & 5.870108 & 2.92$\times$2.20 &  17.0 &  35.3 & 0922+364  & 141.466047268(6.73) &  36.2099096(118.2)\\
NGC 2992 & 5.869068 & 6.68$\times$2.52 &$-$13.7 &  105.1 & 0938$-$133  & 145.260622829(6.07) &$-$13.5974958(193.9)\\
NGC 3081 & 5.874081 & 6.80$\times$3.23 & $-$7.0 &  43.7 & 1004$-$217  & 151.693390348(4.80) &$-$21.9890028(109.5)\\
NGC 3089 & 5.873428 & 7.91$\times$3.36 &   2.9 &  60.1 & 1008$-$285  & 152.772989020(7.74) &$-$28.7945604(246.8)\\
NGC 3079 & 5.887221 & 3.31$\times$1.78 &  17.9 &  441.6 & 0954+556  & 149.40910265(63.41) &	55.38271342(38.3)\\
NGC 3227 & 5.874904 & 5.13$\times$1.77 &$-$11.9 &  48.3 & 1022+194  & 156.186706652(2.61) &	19.20567097(41.3) \\
NGC 3786 & 5.878159 & 3.25$\times$1.70 &  21.6 &  47.9 & 1133+344  & 174.113933109(8.91) &	34.1276345(181.3) \\
NGC 4151 & 5.855409 & 3.13$\times$1.78 &  42.4 &  175.1 & 1204+399  & 181.654389055(8.75) &	39.6843742(132.2)  \\
NGC 4180 & 5.868327 & 6.92$\times$1.55 & $-$4.2 &  76.3 & 1212+087  & 183.749638201(8.64)	 &	 8.4895883(191.2)\\
NGC 4235 & 5.906512 & 3.36$\times$1.48 & $-$7.6 &  106.4 & 1212+087  & 183.749638201(8.64)	 &   8.4895883(191.2)\\
NGC 4388 & 5.868400 & 2.76$\times$1.60 &  13.0 &  107.9 & 1222+131  & 186.265597247(4.1) &	12.8869831(138)\\
NGC 4593 & 5.869410 & 6.93$\times$1.52 & $-$6.7 &  145.3 & 1245$-$062  & 192.095731923(4.98) &	$-$6.5360605(154.7)\\
NGC 5290 & 5.883406 & 3.69$\times$2.02 &$-$19.9 &  61.8 & 1357+404  & 209.908726091(7.73) &	40.1939585(120.2)\\
NGC 5506 & 5.872330 & 5.36$\times$1.78 &$-$16.9 &  352.9 & 1402$-$012  & 211.191231047(5.89) &	$-$1.50609643(93.6)\\
NGC 5899 & 5.869410 & 6.93$\times$1.52 & $-$6.7 &  42.1 & 1505+428  & 226.721007704(6.32)	 &	42.65639874(79) \\
NGC 6814 & 5.873225 & 5.88$\times$2.08 &$-$11.6 &  50.0 & 1937$-$101  & 294.988569046(4.17) &$-$10.04486683(98.1)\\
NGC 7314 & 5.876574 & 7.68$\times$3.73 &   3.9 &  81.5 & 2240$-$260  & 340.860036591(6.44) &$-$25.7418576(189.9)\\
NGC 7378 & 5.865883 & 6.82$\times$2.38 &$-$12.3 &  46.7 & 2243$-$123  & 341.575966543(2.11) &$-$12.11424378(33.5)\\
NGC 7465 & 5.876460 & 4.30$\times$2.13 &$-$15.9 &  47.4 & 2258+166  & 345.179129648(7.01) &	16.9206644(148.4)\\
NGC 7479 & 5.872223 & 4.43$\times$2.04 &$-$17.1 &  43.5 & 2307+106  & 347.618823901(2.5)	 &	10.92519353(41.8)\\
\enddata
\caption{VLBA Observations}
\end{deluxetable*} 

\begin{figure*}
\centering
\includegraphics[width=0.9\textwidth]{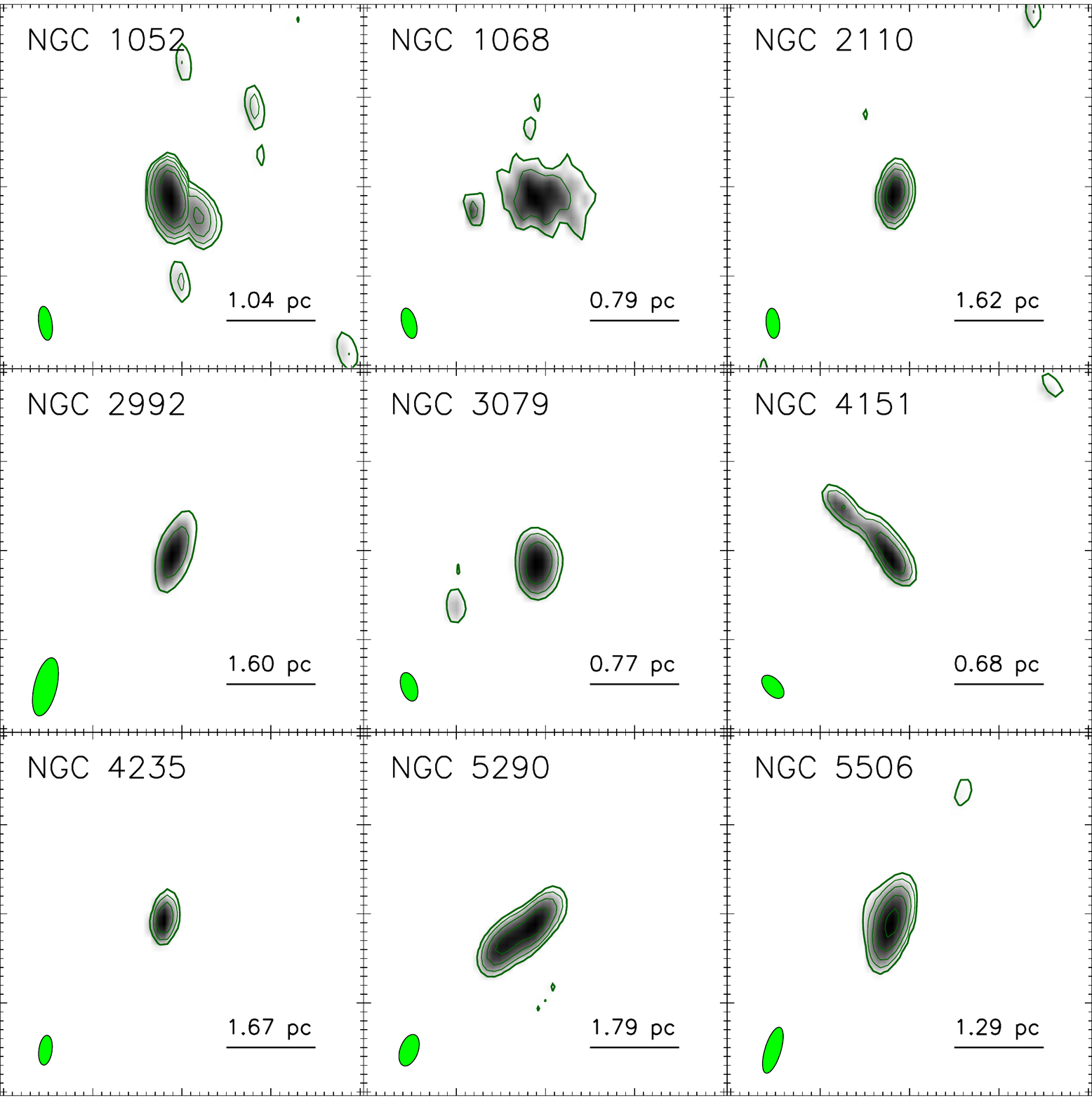}
\caption{5\,GHz (C-band) parsec-scale radio morphologies for the VLBA detected sources in our sample. 
Thick, outer contours represent a 3$\sigma$ flux limit above RMS, with interior contours increasing in powers
of  $3\sigma\times2^n$. Green ellipses to the lower left of each flux distribution represent the synthesized 
beam size for that observation. Scale bars represent an angular size of 10 mas.}
\label{fig:vlbasources}
\end{figure*}

As one of the primary goals of these VLBA snapshot observations is to produce high fidelity images, we devised an observing strategy to maximize $uv$ sky coverage by dividing the 25 targets into bins of similar right ascension.  This produced six groups, each containing three to five sources. We observed each target in a given bin for 20 minutes before slewing to the next target in the bin.  We continued in this fashion, cycling back through each of the targets three times to reach a full, one hour on-source total integration time. In this manner, we were able to track parallactic angles three to five times longer than if we observed each source independently, thereby producing a better-sampled $uv$ plane.  We used phase referencing in order to accurately constrain the phases and ultimately the positions of each target.  Table~\ref{tab:radio_obs} lists the observing parameters including, the center frequency, restoring beam parameters as measured in the plane of the sky and then rotated by the beam position angle, RMS noise of the cleaned image, and the phase calibrator name and position. For each observation, we used a data rate of 2048 Mbps with 2 bit sampling in a single right circular polarization with four intermediate frequency (IF) windows each with a width of 128 MHz and 512 channels.  This recording setup permitted a bandwidth smearing limit of $\sim22\arcsec$ and a time smearing limit of $\sim5\arcsec$. The positions for sources are known to within the error of the positions of the phase reference calibrators, which have all come from the ICRF3 \citep[Charlot et al., in prep.][]{} and have a median formal error of $\sigma_{\alpha} = 6.4$ $\mu$as, $\sigma_{\delta} = 137.3$ $\mu$as.

\begin{deluxetable*}{l r r D{,}{\pm}{4.4} D{,}{\pm}{4.4} c D{,}{\pm}{2.4}} \label{tab:vlbadata}
\tablehead{\colhead{Name} & \colhead{R.A.} & \colhead{Decl.}	& \colhead{F$_{peak}$}  & \colhead{Log\,F$_{peak}^a$} & \colhead{Log\,L$_{peak}$} & \colhead{S$_{int}^a$}  \\  [-0.2cm]
&\colhead{(deg)} & \colhead{(deg)} & \colhead{(mJy bm$^{-1}$)}	& \colhead{($\times10^{-16}$\,erg\,s$^{-1}$\,cm$^{-2}$)} 	& \colhead{(erg s$^{-1})$} &   \colhead{(mJy)}}
\startdata
NGC 1052 & 40.2699939 &$-$8.2557642 & 745,15      & 437,8.8        & 39.38    & 1269,14         \\
NGC 1068 & 40.6696212 &$-$0.0133181 & 0.198,0.016 & 0.116,0.0094   & 35.57    & 2.37,0.19       \\
NGC 2110 & 88.0474013 &$-$7.4562554 &6.89,0.07   & 4.04,0.041     & 37.73    & 11.15,0.17      \\
NGC 2992 & 146.4247680 &$-$14.3262771 &1.01,0.04   & 0.592,0.023    & 36.89    & 1.29,0.10       \\
NGC 3079 & 150.4908485 & 55.6797891 &9.95,0.29   & 5.84,0.17      & 37.24    & 27.02,0.96      \\
NGC 4151 & 182.6357579 & 39.4058501 &2.94,0.14   & 1.73,0.082     & 36.62    & 11.61,0.66      \\
NGC 4235 & 184.2911738 & 7.1915751 &2.56,0.023  & 1.50,0.013     & 37.33    & 3.11,0.049      \\
NGC 5290 & 206.3298364 & 41.7123508 &2.32,0.051  & 1.36,0.029     & 37.35    & 6.49,0.19       \\
NGC 5506 & 213.3119889 &$-$3.2076829 &21.90,0.14  & 12.9,0.0082    & 38.04    & 29.67,0.32      \\
\enddata
$^{a}$Peak flux values, corresponding to the observations in Table~\ref{tab:radio_obs}, derived from CASA's 2-D Gaussian model fitting algorithm and integrated flux densities are determined from flux measured within the 3$\sigma$ outer contour shown in Figure \ref{fig:vlbasources}.
\caption{\centering{VLBA Measurements}}
\end{deluxetable*}


\subsubsection{Data Calibration}
We calibrated our VLBA data using the National Radio Astronomy Observatory (NRAO) Astronomical Image Processing System \citep[\textsc{aips};][]{1996ASPC..101...37V}, release 31DEC19. 
We loaded the data with a calibration (CL) table interval of 0.1 minutes. We loaded in each galaxy with its corresponding phase calibrator independently from the other sources and phase calibrators in the datasets such that all calibrations could be done to one target-phase calibrator pair at a time. We performed all calibrations to the phase calibrator. Using the {\sc vlbautil} module, we corrected for the following: ionosphere delay, Earth orientation, correlator sampler threshold errors, instrument delay, bandpass, amplitude, and parallactic angle. Next, we flagged the data on both the target and the phase calibrator, one source at a time using the {\sc editr} task to remove any bad data and radio frequency interference (RFI). Then, we 
solved for complex amplitudes and phases for the phase calibrator using {\sc{fringe}} and applied the solutions from the phase calibrator to the source by using a two-point interpolation. The calibrations were then applied using the task {\sc{split}}. After these standard AIPS calibration procedures were completed, we moved onto the imaging process. 

\subsubsection{Imaging}
To image each target, we used the \textsc{aips} imaging task {\sc imagr}. For each target we used similar input parameters including a Briggs weighting with \texttt{robust}~$= 5.0$ for a natural weighting in order to maximize sensitivity and a pixel size of 0.8 mas, which Nyquist-samples the synthesized beam. 
We took extreme care in determining the RMS of our VLBA data by initially determining the RMS from a ``dirty image,'' which is made when the number of iterations is set to zero for an initial iteration of  {\sc clean}.  Next, we deconvolved the point spread function using no more than 500 iterations of {\sc clean} to derive the best estimate for the true RMS noise of each image (see Table \ref{tab:radio_obs}). We interactively placed clean boxes around all point source emission and interactively cleaned until a thermal noise limited image was made. For several of our observations, there were significant amounts of RFI and some antennas required complete flagging altogether.  For example, for all sources, the Kitt Peak VLBA antenna suffered a pointing error due to a software bug and thus, required flagging altogether for the duration of our observations.  In the observing session that included NGC 4151, NGC 4180, NGC 4235, NGC 4388, and NGC 4593, we had to flag an additional two antennas due to RFI and other issues in the data.  For these reasons, the RMS values for these targets presented in Table \ref{tab:radio_obs} are above the theoretical RMS by factors of ~4-5.

For the imaging procedure, we started with a field of view of 512 pixels per side, which we increased systematically to search for point source emission, while being mindful not to introduce effects from bandwidth or time smearing. For sources where we did not detect a point source, we imaged out to a radius of $\sim$1$\farcs5$ from the phase center of the source positions.  One of our detected targets, NGC~5290, had a position offset of $+$0$\farcs$35, $-$0$\farcs$2 in R.A.\ and Decl., respectively, from the a priori position as determined from the Pan-STARRS $y$-band data.  Thus, we applied this R.A.\ and Decl.\ offset in the imaging process in order to center the source in the field of view.

Once we had produced initial images, we then proceeded with self-calibration for the nine objects that we detected.  Our self-calibration procedure utilized the \textsc{AIPS} task {\sc calib} and we started by self-calibrating on phase using the initial input image.  Self-calibration is only viable for high S/N detections and thus, we are only able to apply this technique to the nine objects for which we achieved an S/N of $>$15.  This is an iterative process, running CALIB and IMAGR to create new images from the preceding calibrated uv data files.  We repeated this process for several cycles, usually between three or four, until there was little to no improvement in the RMS noise.  Once this was achieved, we then continued self-calibration by calibrating both the amplitudes and phases.  After a few more iterations, our images converged and the RMS noise in the final resulting image no longer improved. Because we utilized phase referencing, our absolute source positions were preserved throughout the self-calibration process. For NGC 1068, however, self-calibration was not applied as this source suffers from severe short spacing artifacts and requires a multi-scale imaging technique. We used four spatial scales for our multi-scale imaging algorithm at 0, 2, 4, and 8 times the restoring beam and we imaged 2048 x 2048 pixels. The final images for our nine detected nuclear AGN sources are presented in Figure \ref{fig:vlbasources}.





\subsection{VLA Observations}
We also include archival images of our sample from the NRAO VLA Archive Survey (NVAS)\footnote{\url{http://archive.nrao.edu/nvas}}, when available, at $\sim4.89$~GHz. 
Largely taken in A-Configuration, we include these datasets in our analysis to compare how peak fluxes, of ostensibly core radio AGN emission, may in fact differ depending on different resolutions. Of the 25 targets in our sample, 19 possessed relevant observations in the archive.

\begin{figure*}
\centering
\includegraphics[width=0.9\textwidth]{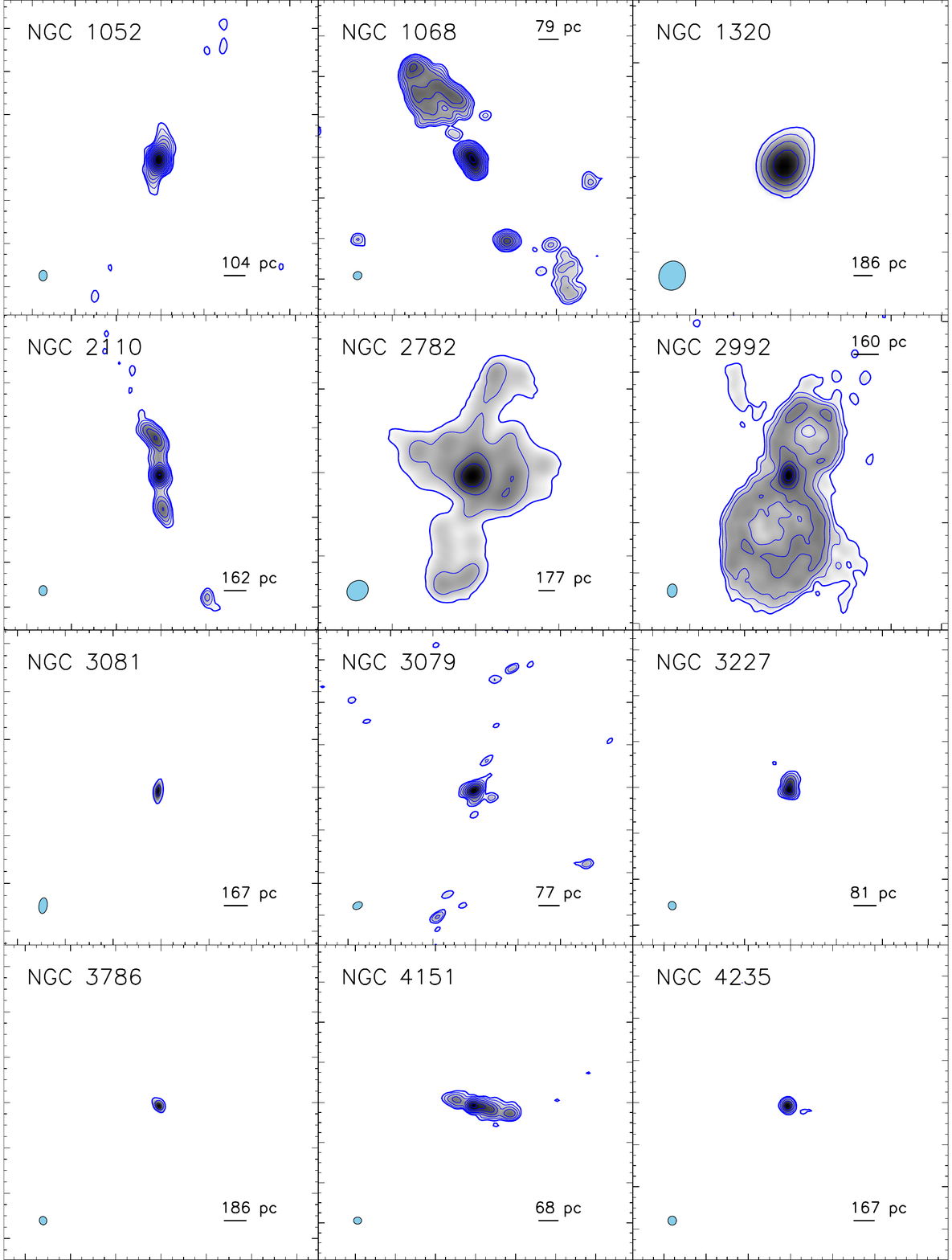}
\caption{C-band (4.9~GHz) hundred-parsec-scale radio morphologies for the VLA observed sources in our sample. 
Thick, outer contours represent a 5$\sigma$ flux limit above RMS, with interior contours increasing in powers 
of $5\sigma\times2^n$. Blue ellipses to the lower left of each flux distribution represent the synthesized beam 
size for that observation. Scale bars represent an angular size of $1\arcsec$.}
\label{fig:vlasources}
\end{figure*}

\begin{figure*}
\ContinuedFloat 
\centering
\includegraphics[width=0.9\textwidth]{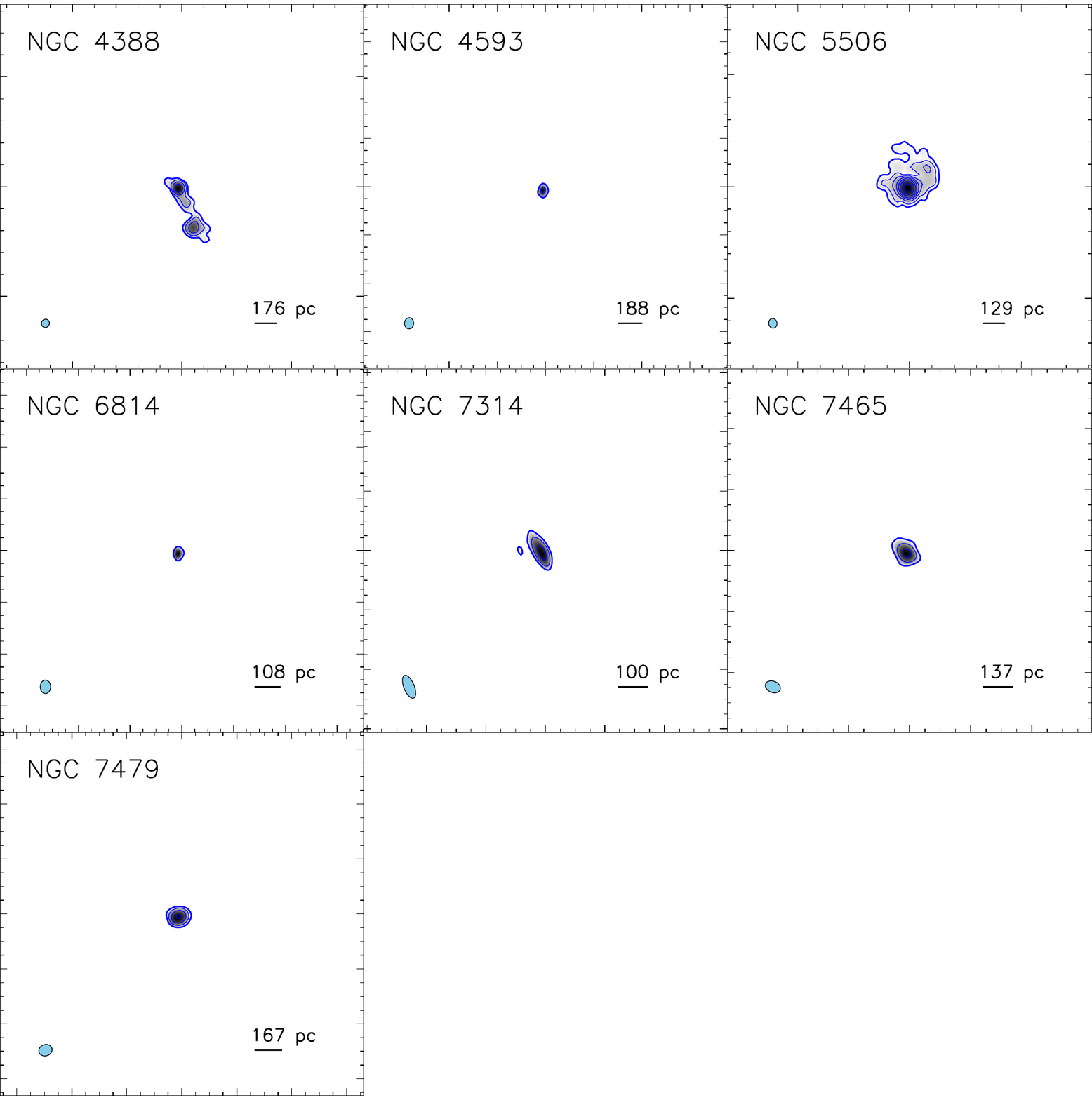}
\caption{{\it (cont.)}}
\label{fig:vlasources_continued}
\end{figure*}

\subsection{Radio Analysis}
The NRAO Common Astronomical Software Applications \citep[\textsc{casa};][]{2007ASPC..376..127M}\footnote{\url{https://casa.nrao.edu}} was used to analyze each image in our VLBA and VLA observations. Using the {\sc viewer} command in \textsc{casa}, we utilized the two-dimensional fitting application to determine both peak flux and integrated flux densities for the 9 detected sources in our VLBA observations and the entirety of the 19 sources observed by the VLA. Peak flux values were measured directly from the image and integrated values were measured by enclosing all nuclear flux greater than 3$\sigma$ or 5$\sigma$ for VLBA and VLA measurements, respectively. The VLA flux distribution in NGC 1068 contains several discrete regions that are not connected to the nuclear emitting region that were also included in its integrated flux density measurement. Table \ref{tab:vlbadata} shows the flux parameters for all nine detections and Table \ref{tab:vladata} shows the peak flux and integrated flux densities for the VLA observations.

\begingroup
\setlength{\tabcolsep}{2pt}
\begin{deluxetable*}{l c c c c c r c c r r} \label{tab:vladata}
\tablehead{\colhead{Name}	& \colhead{Frequency} & \colhead{VLA} & \colhead{Restoring Beam} & \colhead{Beam Angle} & \colhead{RMS} & \colhead{F$_{peak}$}  & \colhead{F$_{peak}$} & \colhead{Log\,L$_{peak}$} & \colhead{S$_{int}^a$} & \colhead{F$_{peak}$ /} \\  [-0.2cm]
& \colhead{(GHz)} & \colhead{Config.} & \colhead{($\alpha \times \delta$; $''$)}& \colhead{(deg)} & \colhead{{\scriptsize($\mu$Jy\,bm$^{-1}$)}} & \colhead{{\scriptsize(mJy bm$^{-1}$)}}	& \colhead{{\scriptsize (erg\,s$^{-1}$\,cm$^{-2}$)}} 	& \colhead{(erg s$^{-1})$} &   \colhead{(mJy)}	& \colhead{VLBA$_{RMS}$}}
\startdata
\multicolumn{3}{l}{\underline{VLBA Detected Sources}} & & & & \\
NGC 1052 & 4.86 & A & 0.50$\times$0.38   &$-$3$\fdg$2    & 177.0 & 1808      & 8.79$\times10^{-14}$  & 39.67   & 1889    & 142.1    \\
NGC 1068 & 4.86 & A & 0.43$\times$0.39   &$-$67$\fdg$1   & 80.6  & 252       & 1.22$\times10^{-14}$  & 38.59   & 980     & 3871.0   \\
NGC 2110 & 4.99 & A & 0.45$\times$0.36   &$-$2$\fdg$6    & 78.9  & 53.5      & 2.67$\times10^{-15}$  & 38.55   & 108     & 110.1   \\
NGC 2992 & 4.86 & A & 0.53$\times$0.39   &$-$5$\fdg$2    & 18.3  & 8.0       & 3.88$\times10^{-16}$  & 37.71   & 57.7    & 25.3   \\
NGC 3079 & 4.99 & A & 0.48$\times$0.33   &$-$63$\fdg$5   & 162.0 & 58.4      & 2.91$\times10^{-15}$  & 37.95   & 89.0    & 44.1   \\
NGC 4151 & 4.99 & A & 0.40$\times$0.35   & 88$\fdg$9    & 90.4  & 38.2      & 1.91$\times10^{-15}$  & 37.66   & 102.0   & 72.8   \\
NGC 4235 & 4.86 & A & 0.43$\times$0.40   & 0$\fdg$4     & 59.7  & 4.6       & 2.25$\times10^{-16}$  & 37.50   & 5.17    & 14.5   \\
NGC 5290 & --   & -- & --                 & --           & --    & --        & --                    & --      & --     & --    \\
NGC 5506 & 4.86 & A & 0.43$\times$0.37   & 12$\fdg$2    & 103.0 & 102.      & 4.96$\times10^{-15}$  & 38.63   & 154     & 96.3   \\
\hline
\multicolumn{3}{l}{\underline{VLBA Non-detections}} & & & & \\ 
NGC 1320 & 4.86 & B & 1.56$\times$1.39   &$-$21$\fdg$8   & 22.9  & 1.65      & 8.04$\times10^{-17}$  & 37.15   & 1.92    & 20.9    \\
NGC 2782 & 4.86 & B & 1.34$\times$1.16   &$-$54$\fdg$4   & 60.3  & 1.50      & 7.29$\times10^{-17}$  & 37.07   & 2.58    & 17.3    \\
IC 2461  & --   & -- & --                 & --           & --    & --        & --                    & --      & --     & --      \\
NGC 3081 & 4.86 & A & 0.66$\times$0.35   &$-$8$\fdg$2    & 54.5  & 0.500     & 2.43$\times10^{-17}$  & 36.54   & 0.502   & 4.5     \\
NGC 3089 & --   & -- & --                 & --           & --    & --        & --                    & --      & --     & --      \\
NGC 3227 & 4.86 & A & 0.37$\times$0.34   & 18$\fdg$2    & 91.9  & 10.8      & 5.24$\times10^{-16}$  & 37.25   & 22.2    & 88.0    \\
NGC 3786 & 4.86 & A & 0.37$\times$0.33   & 14$\fdg$8    & 87.6  & 1.70      & 8.26$\times10^{-17}$. & 37.16   & 2.51    & 13.8    \\
NGC 4180 & --   & -- & --                 & --           & --    & --        & --                    & --      & --     & --      \\
NGC 4388 & 4.86 & A & 0.38$\times$0.36   &$-$28$\fdg$5   & 44.8  & 3.17      & 1.54$\times10^{-16}$  & 37.38   & 9.26    & 19.7    \\
NGC 4593 & 4.86 & A & 0.46$\times$0.38   &$-$2$\fdg$0    & 94.0  & 1.15      & 5.59$\times10^{-17}$  & 37.00   & 1.23    & 10.3    \\
NGC 5899 & --   & -- & --                 & --           & --    & --        & --                    & --      & --     & --      \\
NGC 6814 & 4.86 & A & 0.52$\times$0.4    &$-$0$\fdg$6    & 135.0 & 1.16      & 5.64$\times10^{-17}$  & 36.52   & 1.40    & 8.9     \\
NGC 7314 & 4.86 & A & 0.82$\times$0.35   & 22$\fdg$0    & 65.9  & 1.72       & 8.36$\times10^{-17}$  & 36.53   & 2.00   & 8.3     \\
NGC 7378 & --   & -- & --                 & --           & --    & --        & --                    & --      & --     & --      \\
NGC 7465 & 4.86 & A & 0.51$\times$0.37   & 66$\fdg$6    & 48.6  & 1.21      & 5.88$\times10^{-17}$  & 36.76   & 1.61    & 10.8    \\
NGC 7479 & 4.86 & A & 0.49$\times$0.41   &$-$72$\fdg$6   & 51.4  & 2.43      & 1.18$\times10^{-16}$  & 37.22   & 2.46    & 18.7    \\
\enddata
\footnotesize{$^{a}5\sigma$ contour}
\caption{VLA archival observations and measurements}

\end{deluxetable*}
\endgroup

\subsection{Swift XRT Observations} \label{subsection: Swift XRT}
We obtained Target of Opportunity (ToO) observations (PI: N.\ Secrest) of 20 out of 25 of our sample with the Neil Gehrels Swift Observatory X-ray Telescope \citep[XRT;][]{2003SPIE.4851.1320B,2004SPIE.5165..217H,2005SSRv..120..165B}, which has a PSF with a half-power diameter of $18\arcsec$ and FWHM of $7\arcsec$ at 1.5~keV, and a positional accuracy of $3\arcsec$. These observations were simultaneous with our VLBA observations to obtain a simultaneous snapshot of the radio and X-ray properties of our volume-complete sample of AGNs, as is generally warranted when comparing the radio and X-ray properties of AGNs \citep[e.g.,][]{2003MNRAS.345.1057M,2009ApJ...706..404G}. Three of the five objects we did not acquire observations of were visibility-constrained and the ToOs for the remaining objects, IC~2451 and NGC~2992, were approved but not executed. We requested 1.8~ks integrations for all ToOs, with one ToO set up in Windowed Timing (WT) mode to avoid possible pileup and the remaining ToOs in Photon Counting (PC) mode following the recommendation of the Science Operations Team. Finally, although our ToO observation for NGC~2992 was not executed, a separate XRT observation of NGC~2992 was taken during the requested time, in PC mode, which we include with our data.

We generated X-ray spectra for our data using the online XRT product generator \citep{2009MNRAS.397.1177E},\footnote{\url{https://www.swift.ac.uk/user_objects/}} allowing the routine to center the X-ray sources within the default $1\arcmin$ search radius. For objects where centroiding failed due to faintness, such as NGC~4180, we turned off centroiding and set the extraction coordinates to the target coordinates used for the ToO, which were the same as the VLBA targeting coordinates.

\begin{deluxetable*}{rcrrcccc}
\tablehead{\colhead{Target} & \colhead{Swift Target ID} & \colhead{R.A.} & \colhead{Decl.} & \colhead{Mode} & \colhead{UTC Start} & \colhead{UTC Stop} & \colhead{Obs. Time} \\ [-0.2cm]
                                             &                     & \colhead{deg}  & \colhead{deg}   &                             &                                &                                   & \colhead{second}}
\startdata
NGC 2110 & 35459 & 88.0473918 &$-$7.45625094 & WT & 4/12/19 17:20 & 4/13/19 00:00 & 1525 \\
NGC 2782 & 37237 & 138.5212787 & 40.11369022 & PC & 4/14/19 23:09 & 4/16/19 07:53 & 1670 \\
NGC 2992 & 35344 & 146.4247756 &$-$14.32626689 & PC & 4/14/19 23:09 & 4/15/19 07:53 & 1505 \\
NGC 3081 & 37244 & 149.8731005 &$-$22.82631476 & PC & 4/14/19 23:09 & 4/15/19 07:53 & 1675 \\
NGC 3089 & 11290 & 149.9028701 &$-$28.33129443 & PC & 4/14/19 23:09 & 4/15/19 07:53 & 1790 \\
NGC 3079 & 37245 & 150.4908469 & 55.67979744 & PC & 4/16/19 23:00 & 4/17/19 04:22 & 1580 \\
NGC 3227 & 31280 & 155.8774015 & 19.86505766 & PC & 4/16/19 23:00 & 4/17/19 04:22 & 1610 \\
NGC 3786 & 80684 & 174.9271391 & 31.90942732 & PC & 4/16/19 23:00 & 4/17/19 04:22 & 1735 \\
NGC 4151 & 34455 & 182.6357547 & 39.40584860 & PC & 4/21/19 21:41 & 4/22/19 06:38 & 1715 \\
NGC 4180 & 36654 & 183.2626924 & 7.03891255 & PC & 4/21/19 21:41 & 4/22/19 06:38 & 1630 \\
NGC 4235 & 11308 & 184.2911678 & 7.19157597 & PC & 4/21/19 21:41 & 4/22/19 06:38 & 1580 \\
NGC 4388 & 35464 & 186.4449188 & 12.66215153 & PC & 4/21/19 00:00 & 4/22/19 00:00 & 1755 \\
NGC 4593 & 37587 & 189.9143400 &$-$5.34417010 & PC & 4/21/19 00:00 & 4/22/19 00:00 & 1670 \\
NGC 5290 & 11388 & 206.3297085 & 41.71241871 & PC & 5/7/19 00:00 & 5/8/19 00:00 & 1275 \\
NGC 5506 & 37274 & 213.3119888 &$-$3.20768334 & PC & 5/7/19 03:45 & 5/7/19 09:42 & 1440 \\
NGC 5899 & 36624 & 228.7634964 & 42.04991289 & PC & 5/7/19 03:45 & 5/7/19 09:42 & 1665 \\
NGC 6814 & 32477 & 295.6690092 &$-$10.32345792 & PC & 5/9/19 09:45 & 5/9/19 18:27 & 1665 \\
NGC 7314 & 81861 & 338.9424567 &$-$26.05043820 & PC & 5/9/19 09:45 & 5/9/19 18:27 & 1605 \\
NGC 7378 & 88600 & 341.9486864 &$-$11.81658744 & PC & 5/9/19 09:45 & 5/9/19 18:27 & 1545 \\
NGC 7465 & 11341 & 345.5039963 & 15.96477472 & PC & 5/9/19 09:45 & 5/9/19 18:27 & 1765 \\
NGC 7479 & 37294 & 346.2359605 & 12.32295297 & PC & 5/9/19 09:45 & 5/9/19 18:27 & 1575
\enddata
\caption{Swift XRT Observations}
\end{deluxetable*}

\subsection{X-ray Analysis} \label{subsec: x-ray analysis}
Similar to the procedure outlined in \citet{2017ApJS..233...17R}, we carried out a joint X-ray spectral analysis by combining the contemporaneous XRT data with the time-averaged 105-month BAT data, and fitting models of increasing complexity. We fit the X-ray spectra using \textsc{xspec} \citep{1996ASPC..101...17A}, version 12.10.1f, using \texttt{phabs} to model both intrinsic and Galactic absorption for consistency with the MYTorus model \citep{2009MNRAS.397.1549M} that we use for Compton-thick sources, which requires the \citet{1989GeCoA..53..197A} abundances and the \citet{1996ApJ...465..487V} photo-ionization cross-sections. For Galactic absorption, we use the Swift online Galactic $N_\mathrm{H}$ tool,\footnote{\url{https://www.swift.ac.uk/analysis/nhtot/}} which uses the method of \citet{2013MNRAS.431..394W}.

We began by fitting a single power-law component \texttt{pow}, allowing the normalizations to vary between the XRT and BAT data to account for variability. If soft X-ray absorption is clearly present, we appended a \texttt{phabs} component to account for intrinsic absorption, again allowing the normalizations to vary. For some sources, a soft excess is present. This we fit with either a black-body component (\texttt{bb}) or an additional, non-absorbed power-law component. For sources in which the best-fit $N_\mathrm{H}$ approached Compton-thick levels ($>10^{24}$~cm$^{-2}$), we used MYTorus. We note that because of the $\sim10$~keV coverage gap between the XRT and BAT data, for some sources we relied on the spectral curvature in the BAT data to determine if a source is Compton-thick \citep[e.g.,][]{2016ApJ...825...85K}. In all cases, the 2--10~keV flux we calculate is rest-frame, corrected for absorption, and from the intrinsic power-law component only. For Compton-thick sources, the scattered component may dominate the high-energy continuum \citep[e.g.,][]{2009MNRAS.397.1549M}, necessitating the use of a physical model such as MYTorus to infer the strength of the intrinsic power-law continuum. In order to constrain the intrinsic flux from the power-law component at the epoch of the XRT data, we freeze the best-fit spectral parameters corresponding to components existing at large physical scales and unlikely to exhibit significant short-term variability, such as a black-body or plasma emission component, or optically-thin scattered emission. We then fit the XRT data separately from the fit that included the BAT data, determining the 90\% upper limit on the power-law normalization allowable by the XRT data. We note that when using the MYTorus model the normalization of the scattered continuum is usually held fixed to that of the intrinsic power-law continuum. Because there are likely time lags of unknown duration between the intrinsic power-law emission and the scattered continuum, it is possible that the intrinsic power-law emission at the epoch of the XRT data may be considerably stronger than the scattered continuum. Fixing the scattered continuum means that a much larger range of power-law emission is allowed by the XRT data, yielding a more conservative upper limit.

As a check to ensure we are performing a core X-ray luminosity to core radio luminosity comparison, we additionally fit the Swift BAT data alone, which removes model assumptions and potential contamination from diffuse X-ray emission at softer energies. We fit the BAT spectra with the simplest model required to adequately fit the data ($\chi^2/\mathrm{dof}<2$), where dof is the degrees of freedom. For 15 out of 25 objects a simple power-law was sufficient; for 5 objects a power-law with a high-energy cutoff was required; for the remainder either a power-law or cutoff power-law with an absorption component was warranted. We do not physically interpret these models or attempt to de-convolve an intrinsic power-law continuum from the fit as the 105-month BAT catalog data are limited to only 8 spectral bins. We appended the \texttt{cflux} convolution component in front of the best-fit spectral model and fit for the total observed 14--195~keV flux after freezing the power-law normalization. We found that we were able to more robustly estimate the uncertainty of the 14--195~keV flux using Markov Chain Monte Carlo (MCMC) with the default Goodman-Weare algorithm with 10 walkers. For every object we ran chains of length $10^5$ after determining the appropriate burn-in period. For NGC~4180, the MCMC chain produced a bimodality in $\Gamma$ with one mode at unphysically low values $<1$. We alleviated this by setting a Gaussian prior on $\Gamma$ with a mean of $2.00$ and standard deviation of $0.38$, taken from the distribution of $\Gamma$ for all AGN in the 105-month catalog, using the \texttt{bayes} command.

Finally, as we are comparing the XRT data to data from other facilities, we include in the X-ray fluxes a systematic uncertainty floor of 10\% \citep{2005SSRv..120..165B}, which we add in quadrature to the formal errors.




\section{Results and Discussion} \label{section: Results and Discussion}
\subsection{A Comparison of VLA and VLBA Fluxes} \label{subsec: vlavlba}
\begin{figure*}
\includegraphics[width=\columnwidth]{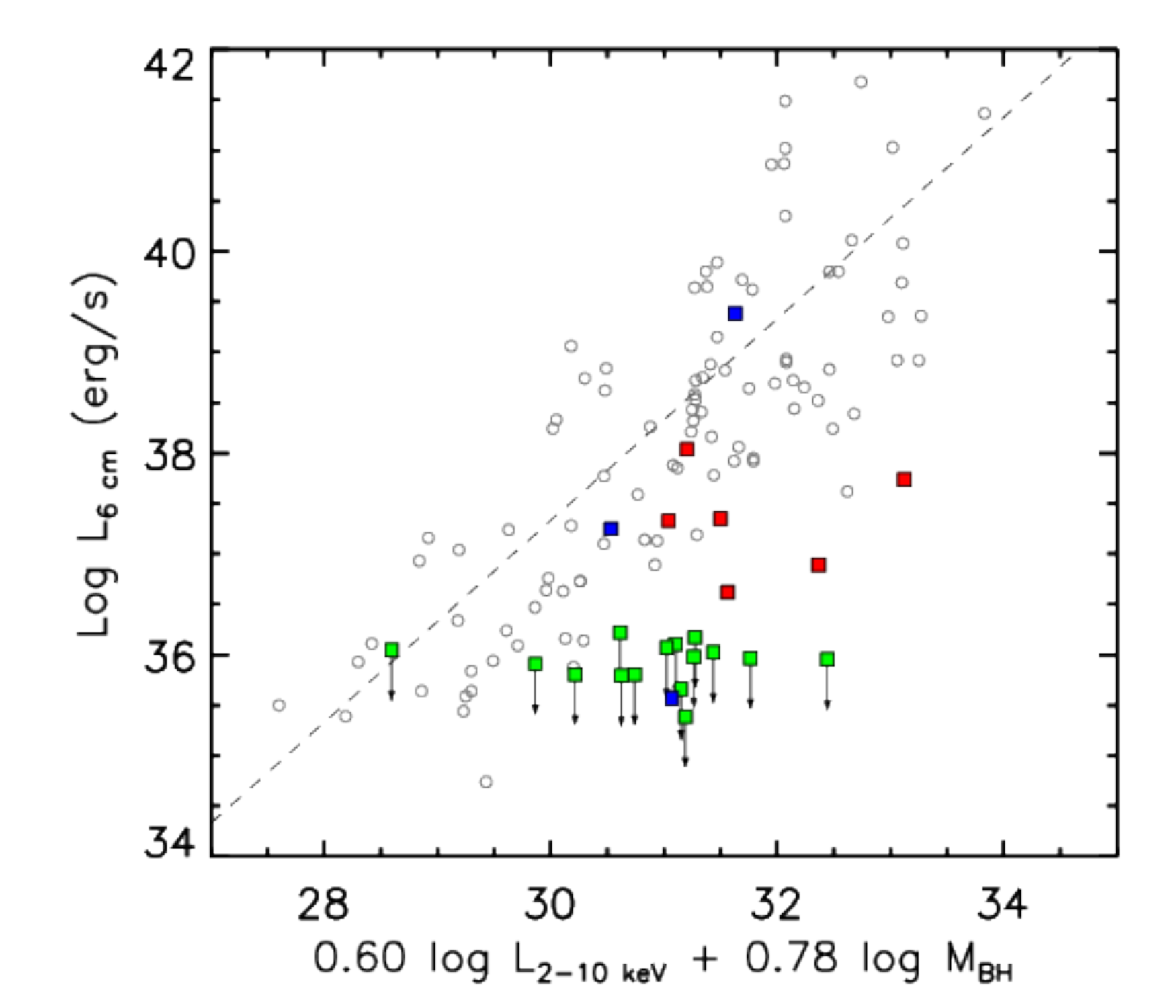}
\includegraphics[width=\columnwidth]{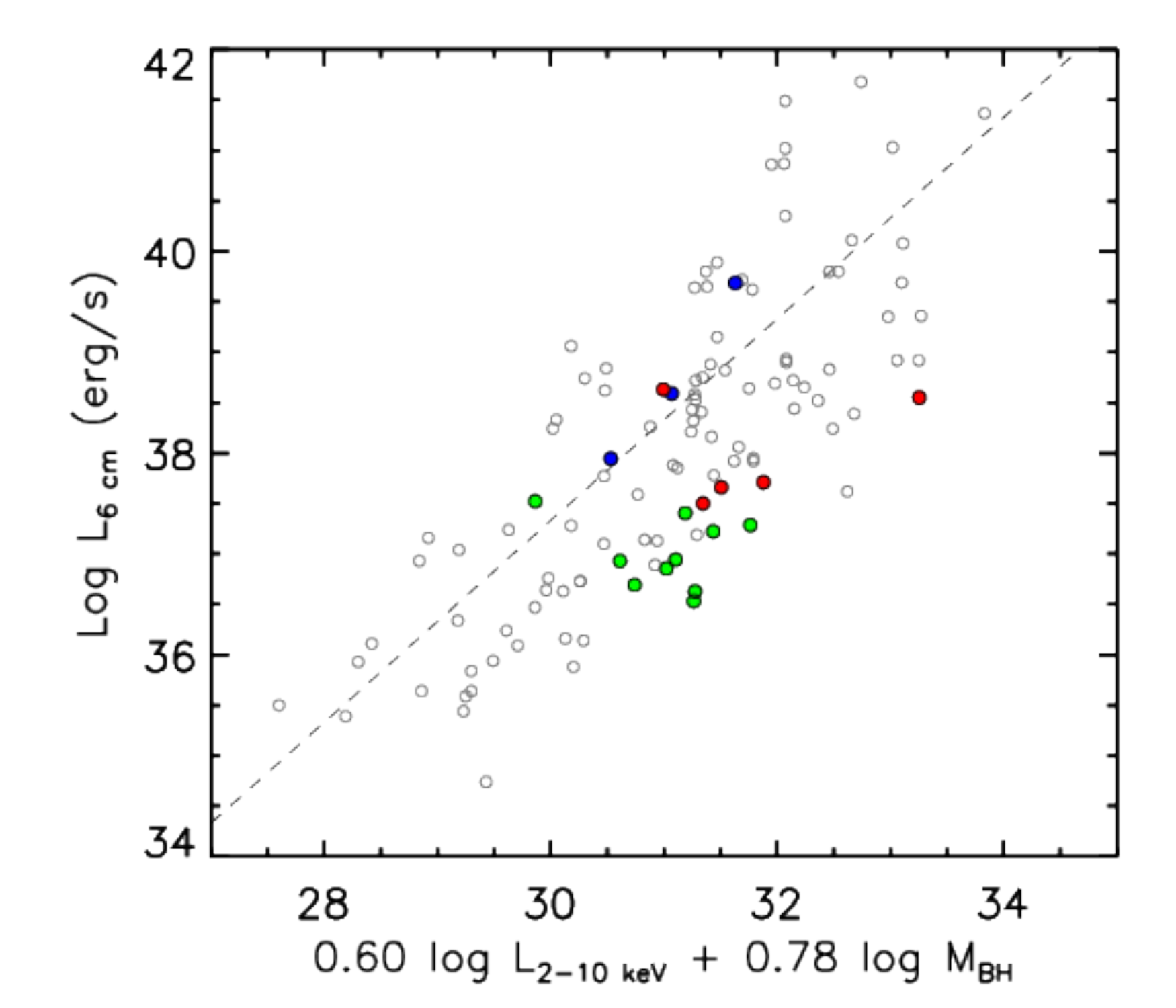}
\caption{Distribution of our sample on the FP described by \citet{2003MNRAS.345.1057M} using peak radio luminosities derived from {\it left:} VLBA observations and {\it right:} VLA observations (when available). Red points represent targets with simultaneous VLBA and Swift XRT observations. Blue points represent targets with new VLBA observations and archival Swift XRT observations. Green points are targets with VLBA observations where we estimate the upper limit of $L_R$ in the targets as $5\times$RMS. Grey circles and dashed line represent measurements and best fit from \citet{2003MNRAS.345.1057M}.}
\label{fig:fundplane}
\end{figure*}

An immediate point of interest from our initial snapshot program was the lack of VLBA detections for a majority of our sample. As our VLBA integration time calculations were derived from the FP for AGNs, we expected detections for the entirety of our sample. The left plot of Figure~\ref{fig:fundplane} places our sample on the FP, in comparison to data from \citet{2003MNRAS.345.1057M}, using peak radio and X-ray luminosities calculated from VLBA and Swift XRT measurements in Tables~\ref{tab:vlbadata} and \ref{tab:XRTdata}, respectively, and black hole mass measurements from Table~\ref{tab:volume}. Targets with VLBA detections are represented by red and blue points which use simultaneous and archival XRT observations, respectively. Green points with arrows mark radio luminosity upper limits for targets with VLBA non-detections, which we define as $3\sigma$ over the RMS in each observation. From this comparison, we find that although several radio luminous targets fall within the \citet{2003MNRAS.345.1057M} distribution, our sample lies largely below the defined plane. As the FP was derived using VLA observations, we replace our peak radio luminosities measured from our VLBA observations with peak radio luminosities from archival VLA observations in Table \ref{tab:vladata}. Using these values, as shown in the right plot of Figure \ref{fig:fundplane}, we find our targets are 100$\%$ detected, versus 36$\%$ with VLBA, and become well-aligned with the FP and the \citet{2003MNRAS.345.1057M} data distribution.

Our findings are consistent with those of \citet{2013MNRAS.432.1138P}, who studied a more nearby ($D_L<22$~Mpc) sample of Seyfert galaxies with the European VLBI Network (EVN), which has a similar maximum baseline to the VLBA ($\sim9000$~km). They find that out of 21 Seyfert galaxies observed with the EVN, 13 (62\%) are detected at 5~GHz. The 90\% confidence interval from binomial statistics for this fraction is 42\%--79\%, whereas for our targets, of which 36\% were detected, the 90\% confidence interval is 20\%--54\%, with a p-value between our study and \citet{2013MNRAS.432.1138P} of $p=0.16$. Nonetheless, there are differences between these samples that are worth noting. First, the \citet{2013MNRAS.432.1138P} sample is again considerably closer, with a mean distance of 14.7~Mpc,\footnote{\citet{2013MNRAS.432.1138P} also use a flat $\Lambda$CDM with $H_0=70$~km~s$^{-1}$~Mpc$^{-1}$ and $\Omega_\mathrm{M}=0.3$.} compared to the mean distance of 30.0~Mpc for our sample. Given the $3\sigma$ luminosity upper limits listed in their Table~1, the mean RMS of their EVN data is $\sim57$~$\mu$Jy, about half as sensitive as our VLBA observations. The difference in mean distance and sensitivity implies that the 5~GHz observations used by \citet{2013MNRAS.432.1138P} are comparable in terms of probing the target luminosities, with upper luminosity limits about $\sim0.6$ times that of our observations. Of the 20 objects in \citet{2013MNRAS.432.1138P} with 2--10~keV X-ray luminosity measurements, the mean X-ray luminosity is $8.1\times10^{41}$~erg~s$^{-1}$, while for our sample the mean luminosity is $1.7\times10^{43}$~erg~s$^{-1}$, indicating that higher X-ray luminosity is not necessarily correlated with greater prevalence of radio core detection at mas scales, given the statistical consistency of the latter between our study and that of \citet{2013MNRAS.432.1138P}. As in this work, \citet{2013MNRAS.432.1138P} also find that the X-ray emission is more correlated with larger-scale radio emission as captured by the archival VLA observations they use, and they suggest that this implies that the hard 2--10~keV luminosity in an AGN is associated with larger-scale structure than the radio core. With both 5~GHz and 1.7~GHz VLBI observations, they are able to estimate spectral index $\alpha$, and find that the disparity between the VLA and VLBI fluxes is less in flatter-spectrum objects, consistent with the emission being truly core-dominated. 

It is interesting to note that other studies aimed at detecting radio quiet quasars on spatial scales of the VLBA using ancillary catalogs such as those from the VLA as a priori source lists find similar detection rates to the ones we find in this work.  For example, the recent work by \citet{Her17} mosaic the COSMOS field with the VLBA at 1.4 GHz and find a detection rate of $\sim$20\% for all VLA detected sources at 3 GHz.  \citet{Her18} followed up this work by combining the Robert C. Byrd Green Bank Telescope together with the VLBA (VLBA+GBT) for increased sensitivity at 1.4 GHz and detected an additional 10 sources that were not previously seen with the VLBA alone.  The overall detection rate for this follow-up study, however, remained at $\sim$20\%.  Similarly, \citet{Mid13} mosaicked the Lockman Hole/XMM field with the VLBA at 1.4 GHz and out of their parent sample of 217 sources, only 65 sources ($\sim$30\%) were, in fact, detected at their 1$\sigma$ sensitivity threshold of $\sim$20 $\mu$Jy beam$^{-1}$.  Although these studies present results at a different frequency than our study (1.4~GHz versus 6~GHz), the radio emission across both of these frequencies is expected to be dominated by synchrotron, non-thermal emission processes \citep[see e.g.,][for a discussion on radio emission in galaxies]{1992ARA&A..30..575C}.  There are clear differences in the observing strategies, sensitivity limits, and populations of radio quasars studied by \citet{2013MNRAS.432.1138P}, \citet{Her17, Her18}, \citet{Mid13} and our volume-complete sample, but the detection rates from all of these studies are within the $\sim$20-40\% range, which may be pointing to a similar process responsible for the core radio emission seen in AGNs at these VLBA spatial scales.

\begin{figure*}
\centering
\includegraphics[width=0.32\textwidth]{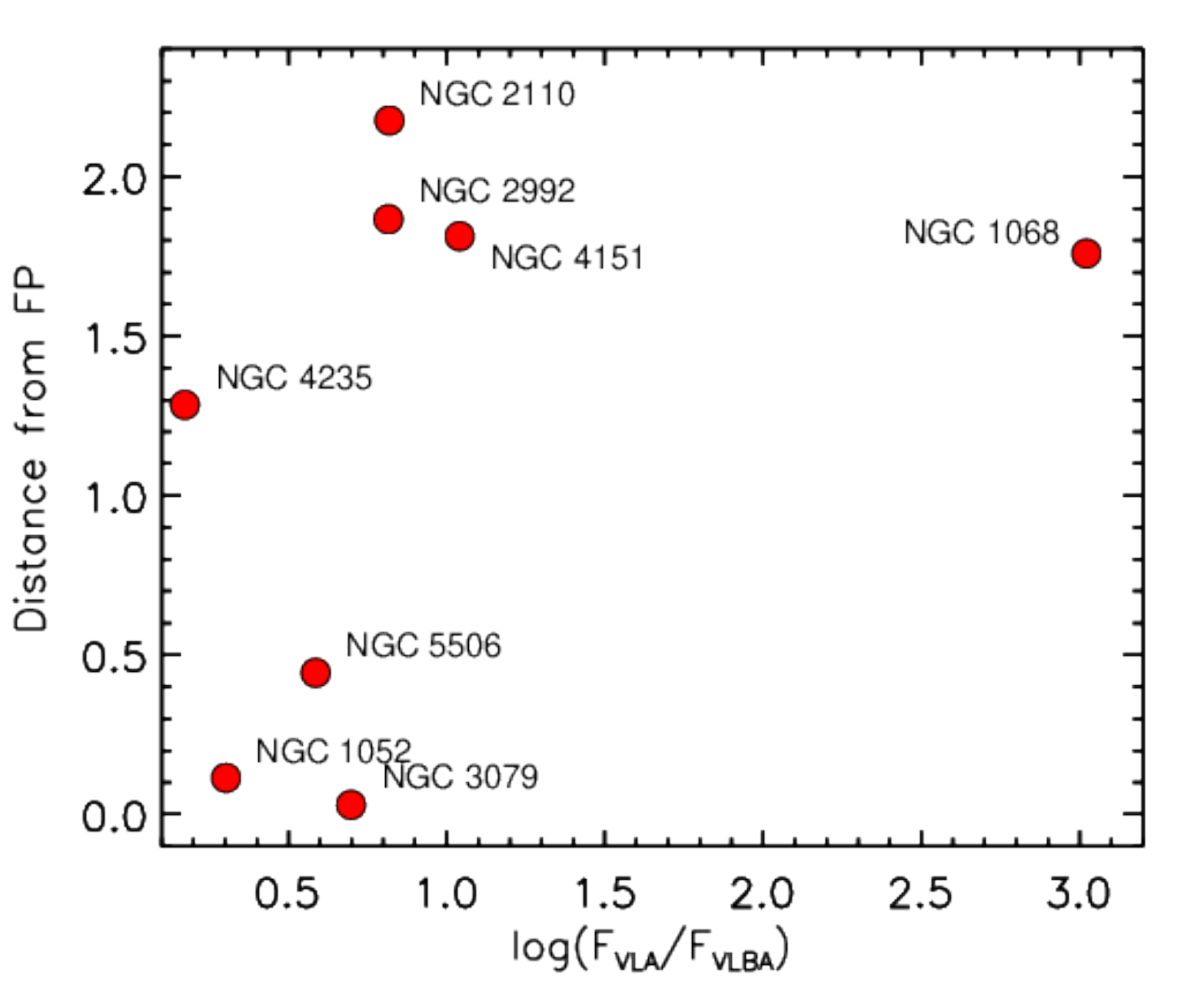}
\includegraphics[width=0.32\textwidth]{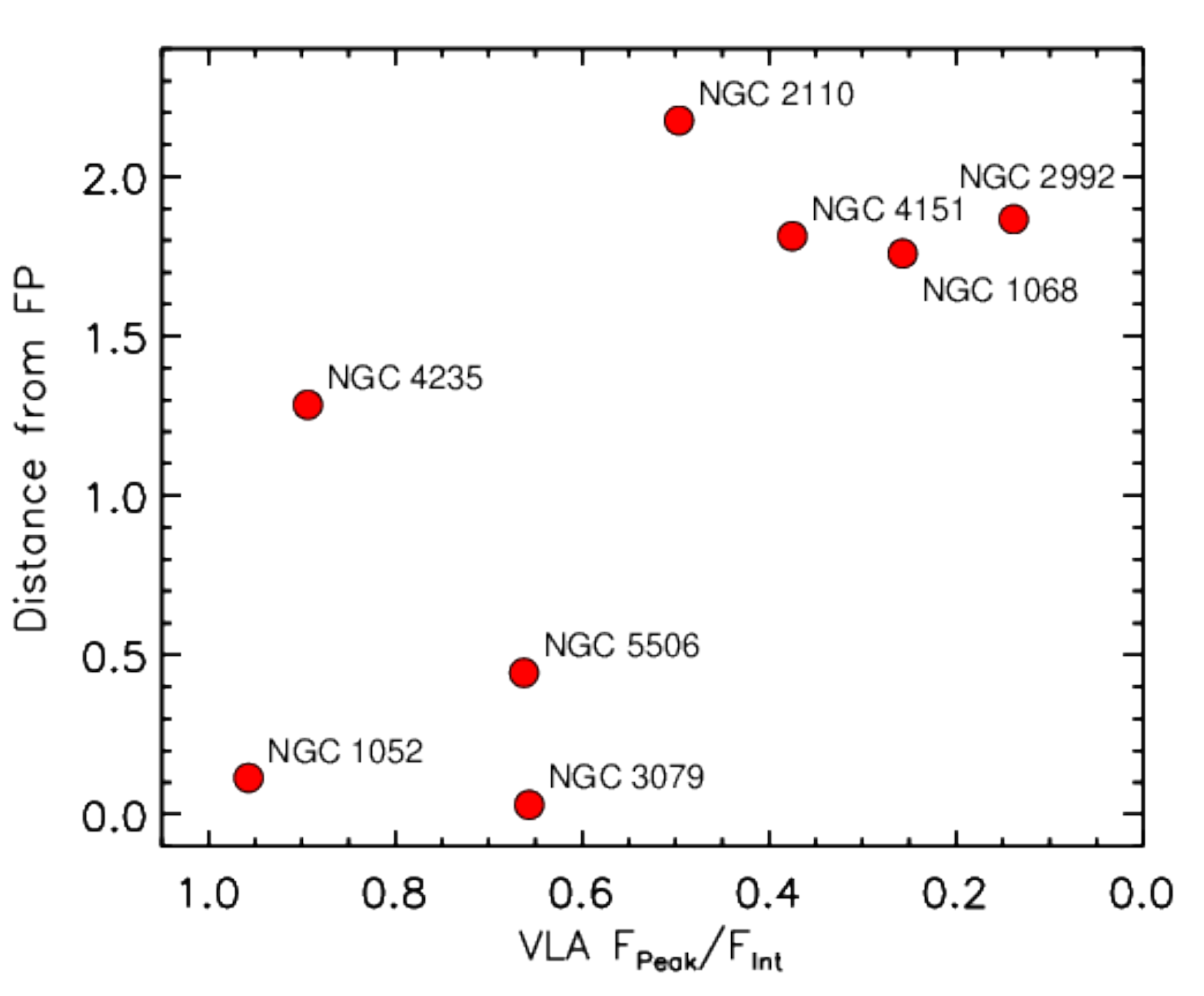}
\includegraphics[width=0.32\textwidth]{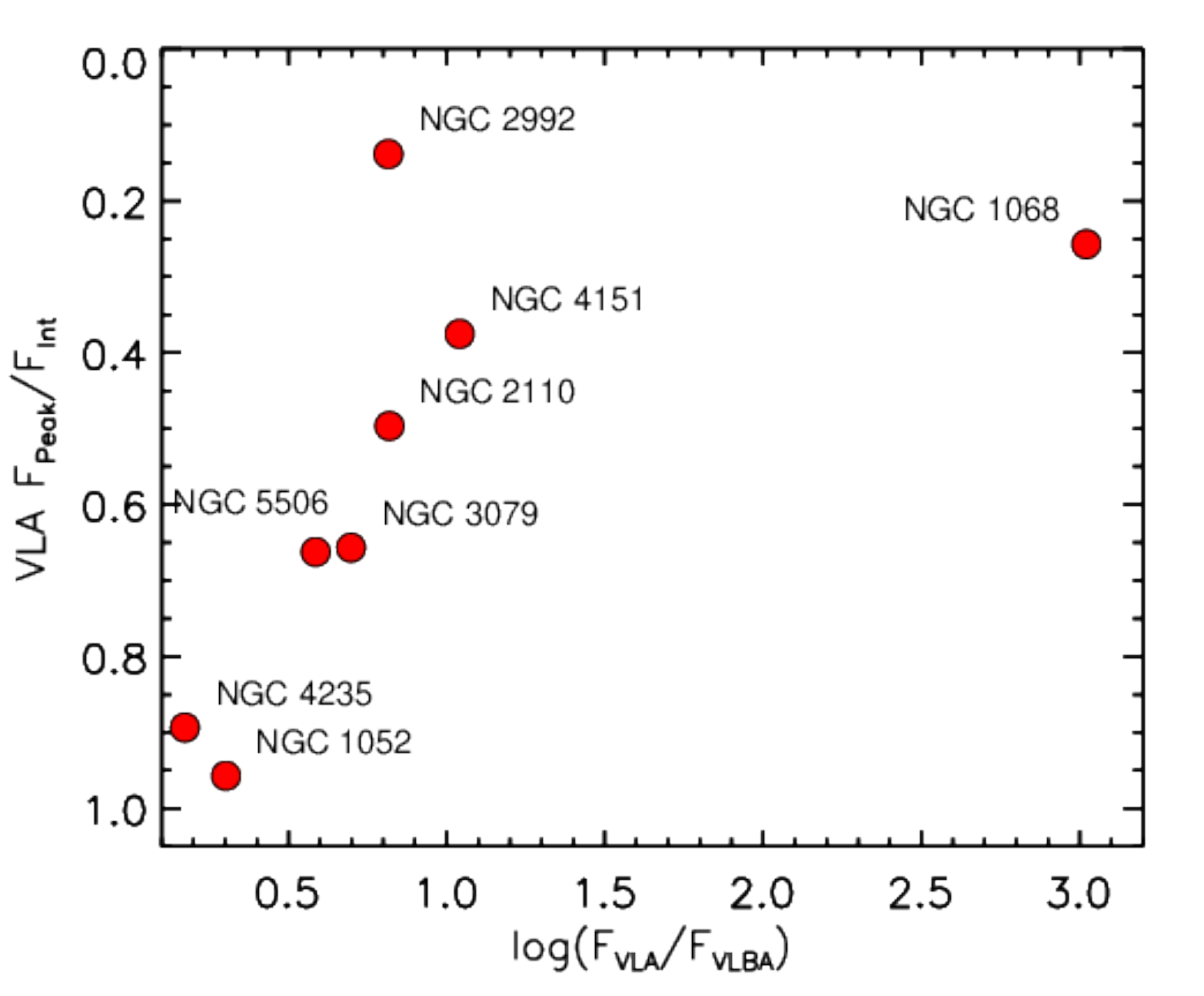}

\caption{Comparisons between ratio of integrated to peak VLA flux, ratio of peak VLA to peak VLBA flux, and distance from the Fundamental Plane in dex.}
\label{fig:ext_vs_rat}
\end{figure*}

To explore this idea further, we compare the VLBA and VLA peak and integrated flux density measurements in our targets. We find the difference in peak flux measurements between VLBA and VLA observations appear to be greater in targets such as NGC~1068 and NGC~2992, with VLBA measurements that place them further from the Fundamental Plane as shown in Figure \ref{fig:ext_vs_rat}.  We also find that targets further from the Fundamental Plane have more extended 6\,cm radio morphologies in our VLA observations. We quantify this by measuring the ratio between the VLA flux peak (F$_{peak}$) and the VLA integrated flux density greater than 5$\times$RMS (F$_{int}$) using values listed in Table \ref{tab:vladata}, where more extended targets have a greater portion of their flux outside the nucleus and a ratio closer to zero. Further, we find a relationship between these two ratios, where AGN with greater VLA/VLBA flux ratios are typically more extended as they exhibit larger VLA F$_{peak}$/F$_{int}$ ratios. Together, these relationships suggest that the emission observed in VLA observations is largely extended, where AGN that deviate further from the Fundamental Plane have larger differences in VLA and VLBA peak flux measurements and exhibit more extended radio emission relative to the nucleus. Additionally, we can show that emission measured in VLA observations is extended in targets with VLBA non-detections by comparing our VLA peak-flux measurements to the VLBA background RMS as listed in Table \ref{tab:vlbadata}. Assuming that the VLA emission originated from a point source, one would expect to detect said source in the VLBA observations with a signal-to-noise $S/N = F_{peak}(VLA) / VLBA_{RMS}$. We measure this hypothetical S/N ratio to be $>$ 3 in all targets with VLA observations, with S/N $>$ 5 in 10 of 11 targets, suggesting a point source would have been detected for each AGN if present. 

Following \citet{Zak14}, we suggest that the extended radio emission observed in the VLA observations for this sample is largely a by-product of relativistic particles accelerated in shocks caused by AGN-driven winds, similar to processes in supernova remnants. Assuming a scenario where radio emission in each of these targets is attributed to a jet emanating from the nucleus, one would expect larger discrepancies between VLA and VLBA observations in compact sources, where the jet is pointed along our line of sight. Instead of being spread across the plane of the sky, the bulk of the extended, contaminating emission in these targets would be contained near the nucleus along our line of sight, captured in VLA observations but resolved out in VLBA observations. As we note, we find larger ratios to exist between VLA and VLBA peak flux measurements in targets with more extended morphologies, thus it is likely that we are not observing an effect dependent on the orientation of the AGN with respect to our line of sight. Instead, in a scenario where emission is formed by shocks via AGN feedback, the observed ratio between VLA and VLBA fluxes would be dependent on the orientation of the AGN with respect to its host galaxy, where targets in our sample with extended VLA morphologies are exhibiting more interactions between the AGN and its host medium than targets with compact VLA morphologies. 

\begin{figure}
\includegraphics[width=\columnwidth]{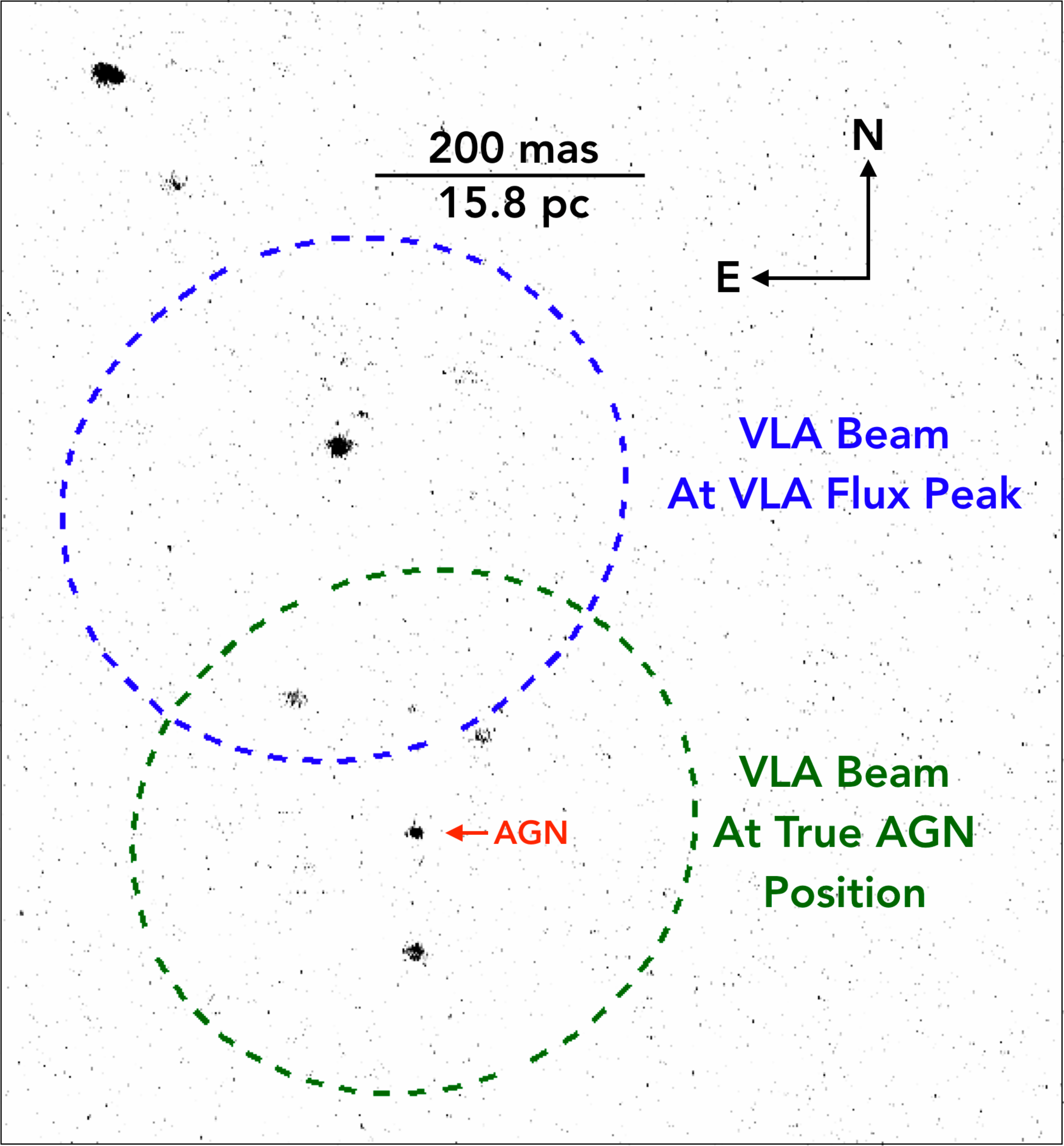}
\caption{Extended 5\,GHz VLBA imaging of NGC~1068 showing the central AGN, and adjacent, extra-nuclear emission. Larger, dashed ellipses represent the synthesized beam size of the corresponding 5\,GHz VLA observation. The northern, blue ellipse is centered on the VLA flux peak, and the southern, green ellipse is centered on the AGN. Both ellipses illustrate that extra-nuclear radio emission can contaminate VLA peak flux measurements. The red arrow points to the nuclear AGN emission shown in Figure \ref{fig:vlbasources}.}
\label{fig:ngc1068comp}
\end{figure}

These interactions translate down to parsec-scale levels, where VLA peak-flux measurements of our AGN are likely contaminated by unresolved extra-nuclear host interactions that resolve out of VLBA peak-flux observations with the AGN becoming much less luminous than hypothesized from the Fundamental Plane relation. We see this explicitly in VLBA observations of NGC 1068 in Figure \ref{fig:ngc1068comp} which exhibit several knots of extra-nuclear emission in addition to the AGN, whose position was previously identified by \citet{Gal04} and references therein. Placing the synthesized beam and flux peak location obtained from the archival VLA observations onto our VLBA observations, we see that the peak flux measured by the VLA is derived largely from extra-nuclear emission north of the nucleus and likely does not contain emission from the AGN itself. While the source at the center of the VLA beam and the central AGN exhibit similar peak fluxes in our VLBA measurements, MERLIN observations from \citet{Gal04} reveal the non-AGN source to be much more luminous at scales slightly larger (MERLIN restoring beam $\sim$60\,mas) than what is resolvable by VLBA (maximum resolvable size $\sim$47\,mas). We note that NGC~1068 is the only source in our sample that exhibits such a misalignment between VLA peak and VLBA peak, which likely accounts for the extreme ratio between VLA and VLBA fluxes for NGC~1068 compared to the seven other sources shown in Figure \ref{fig:ext_vs_rat}. However, even without a misalignment where the VLA peak flux would have been centered over the AGN, Figure \ref{fig:ngc1068comp} also exhibits several knots of extra-nuclear emission adjacent to the central engine which would contaminate VLA peak-flux measurements for NGC~1068.

\begin{figure*}
\includegraphics[width=0.5\textwidth]{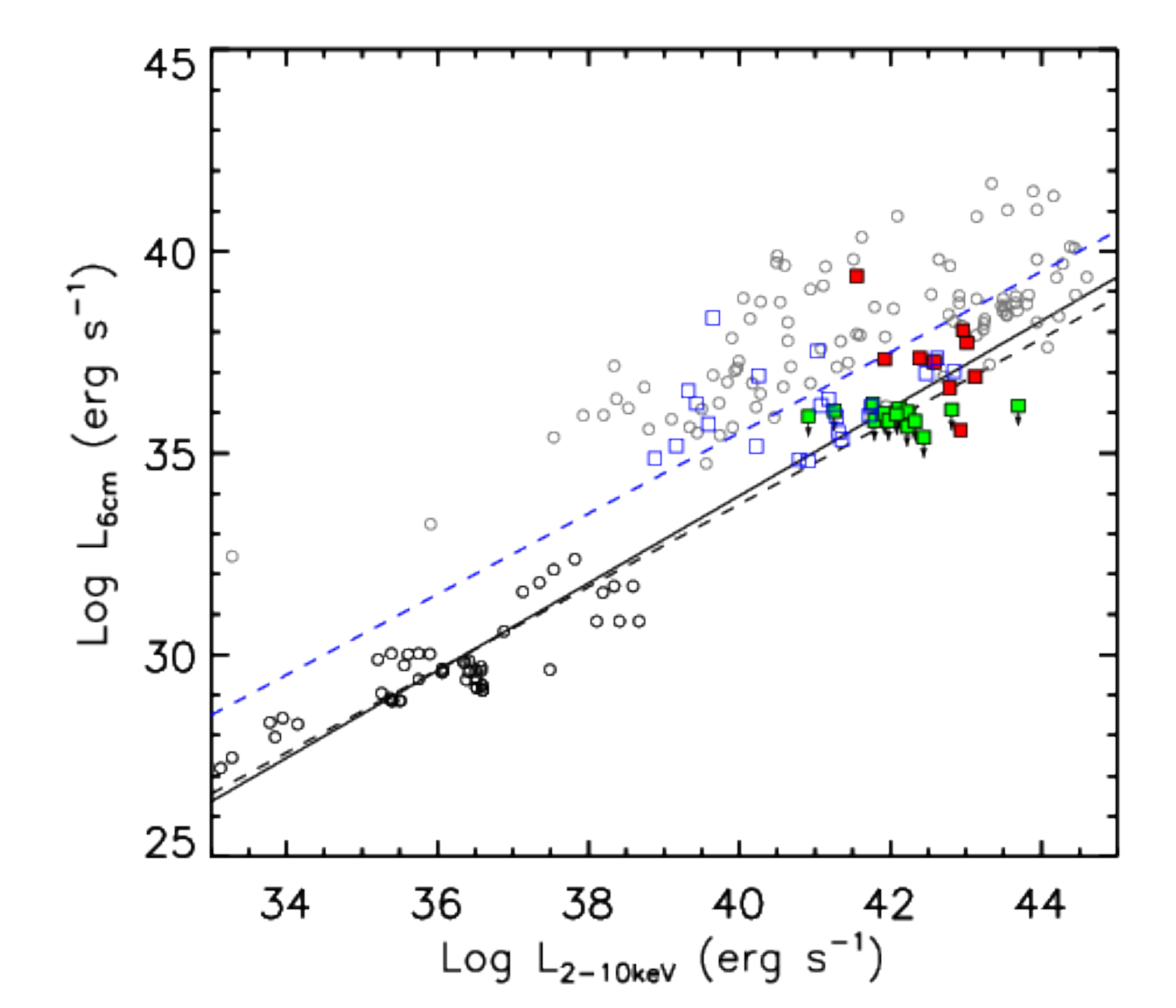}
\includegraphics[width=0.5\textwidth]{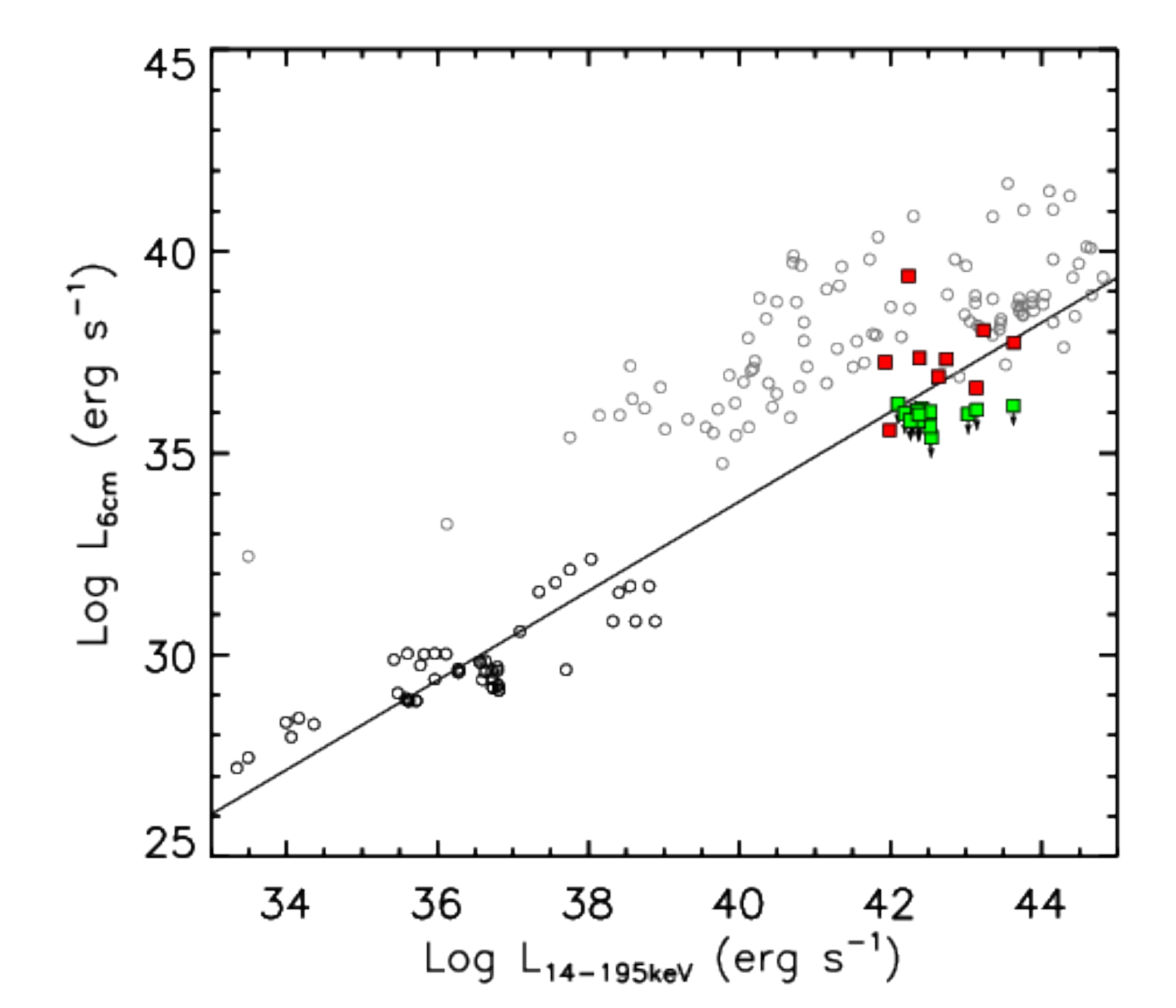}
\caption{Comparison between radio and X-ray luminosities in our sample, \citet{2013MNRAS.432.1138P}, and \citet{2003MNRAS.345.1057M} 
for Galactic and extragalactic black holes. {\it Left:} Red and green squares designate measurements for our detected sources and upper 
limits on undetected sources, respectively. Open blue squares designate measurements from \citet{2013MNRAS.432.1138P}.
Open grey and black circles represent SMBH and GBH measurements from \citet{2003MNRAS.345.1057M}, respectively. Solid line represents the best fit between our detected sources and GBHs from Merloni et al. The black dashed line represents the best fit also including the upper limits from our non-detections. The blue dashed line represents the X-ray radio loudness parameter of $R_{X} \equiv L_{R}/L_{X} = -4.5$.{\it Right:} Similar comparison using 14-195 keV measurements of our sample compared to 2-10 keV luminosities from Merloni et al. scaled by 1.6.}
\label{fig:rvsx}
\end{figure*}

\begin{figure*}
\includegraphics[width=0.5\textwidth]{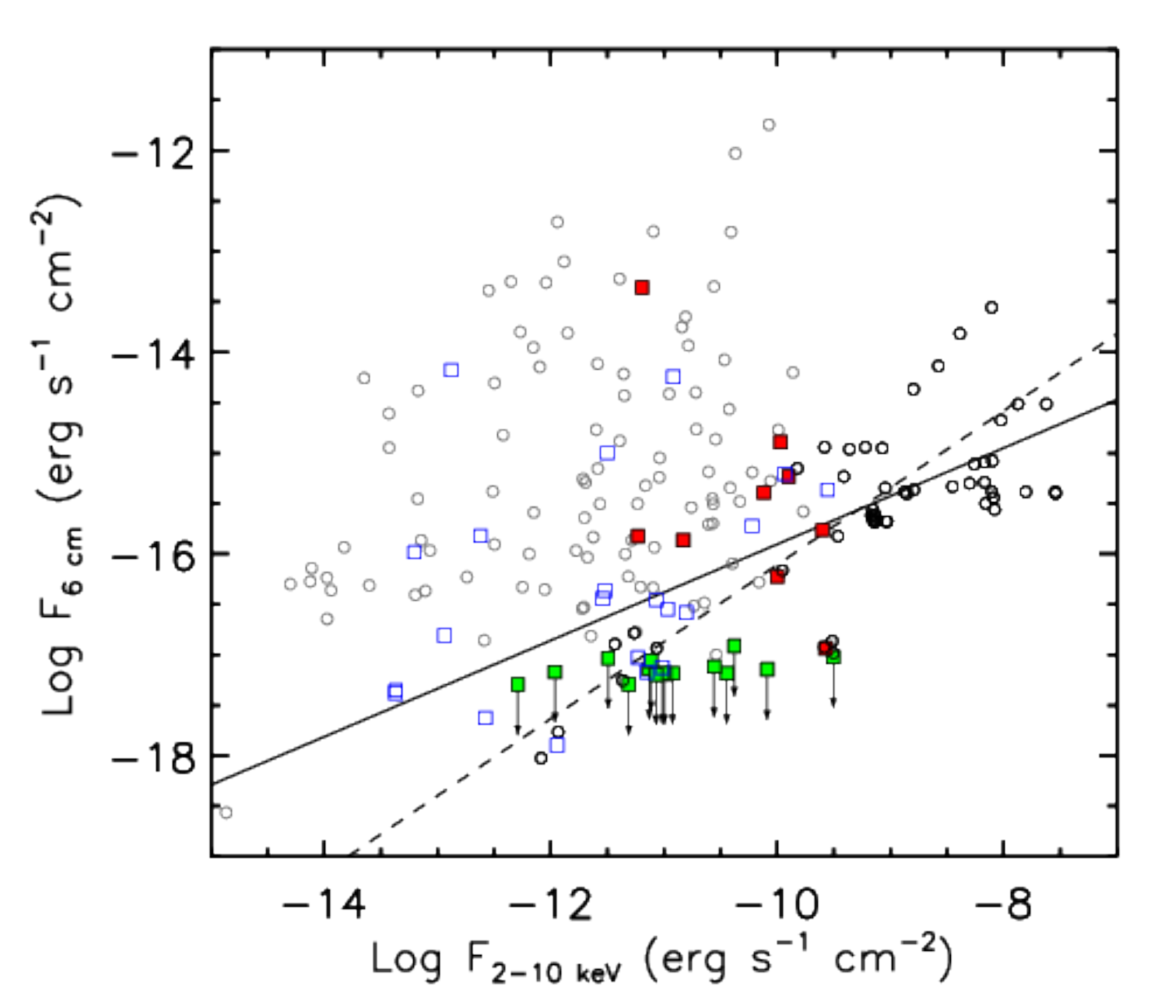}
\includegraphics[width=0.5\textwidth]{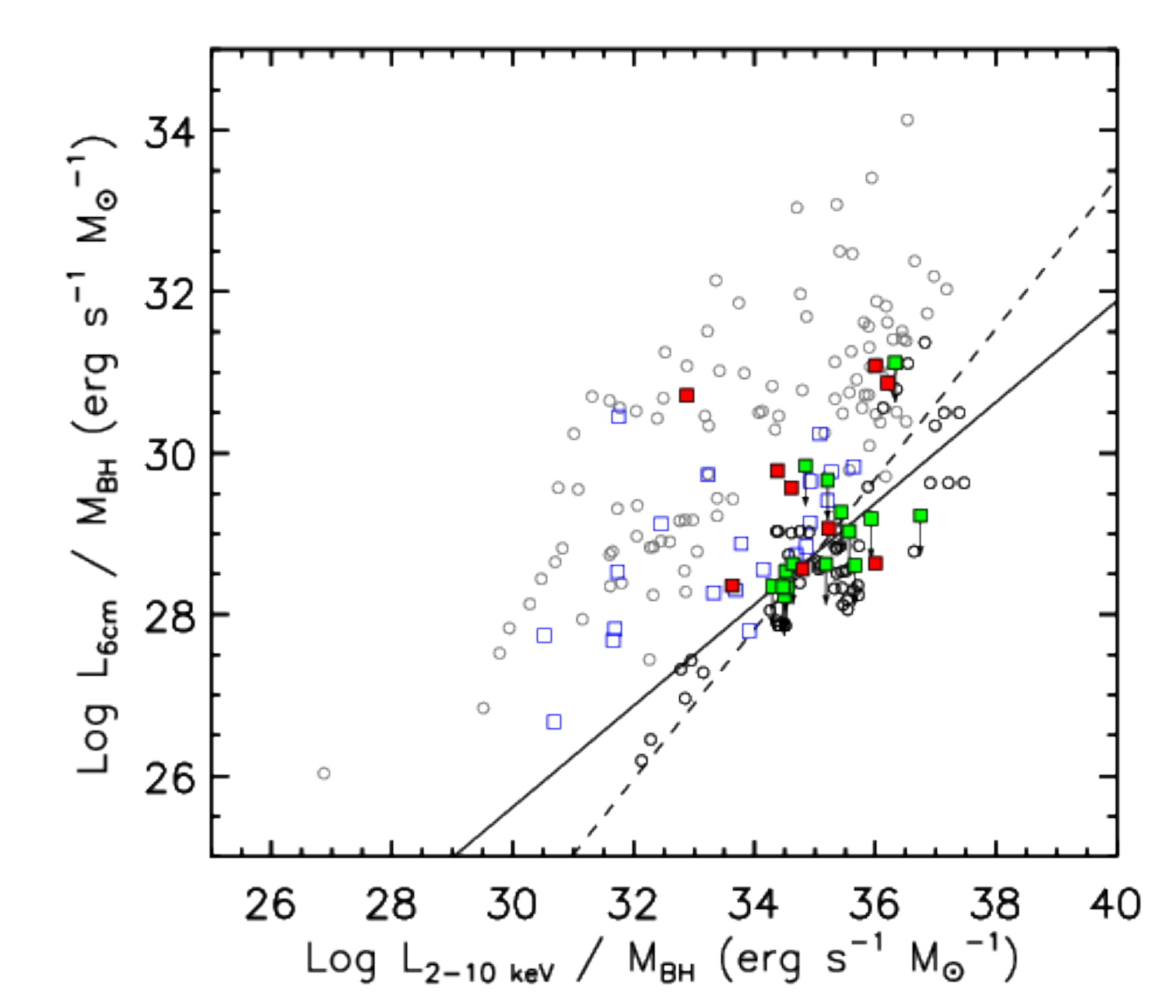}
\caption{Comparison between radio and X-ray flux measurements ({\it left}) and radio and X-ray luminosities divided by black hole mass ({\it right}) 
in our sample and \citet{2003MNRAS.345.1057M} for Galactic and extragalactic black holes. Red and green squares designate measurements for our detected 
sources and upper limits on undetected sources, respectively. Open blue squares designate measurements from \citet{2013MNRAS.432.1138P}. 
Open grey and black circles represent SMBH and GBH measurements from \citet{2003MNRAS.345.1057M}, respectively. Solid line represents the 
best fit between our detected sources and GBHs from Merloni et al. Dashed line represents the best fit also including the upper limits from 
our non-detections.}
\label{fig:rvsxflux}
\end{figure*}

\begin{deluxetable*}{clcrcc}
\tablehead{\colhead{Object} & \colhead{Model} & \colhead{Total stat/dof} & \colhead{$N_\mathrm{H}$} & \colhead{$\Gamma$} & \colhead{$\log_{10}(F_\mathrm{2-10~keV})$} \\ [-0.1cm]
 & & & \colhead{cm$^{-2}$} & & [erg cm$^{-2}$ s$^{-1}$] 
}
\startdata
NGC 2110 & \texttt{phabs*pow + const*pow} & 383.25/473 & $6.8_{-1.7}^{+2.1}\times10^{22}$ & $2.5_{-0.5}^{+0.6}$ & $-10.12_{-0.04}^{+0.05}$ \\
NGC 2782 & \texttt{apec + zpow + const*pow} & 29.31/30 & $1.4_{-0.1}^{+0.3}\times10^{23}$ & $1.8_{-0.5}^{+0.5}$ & $-11.9_{-0.4}^{+0.5}$ \\
NGC 2992 & \texttt{phabs*pow} & 358.22/416 & $7.5_{-1.7}^{+1.9}\times10^{21}$ & $1.6_{-0.2}^{+0.2}$ & $-10.00_{-0.02}^{+0.02}$ \\
NGC 3079 & \texttt{vapec + MYTorus} & 120.30/95 & $8.5_{-1.0}^{+1.2}\times10^{24}$ & $2.2_{-0.1}^{+0.1}$ & $-9.9$ \\
NGC 3081 & \texttt{apec + MYTorus} & 41.05/34 & $6.1_{-0.3}^{+0.4}\times10^{23}$ & $1.96_{-0.01}^{+0.01}$ & $-10.5_{-0.2}^{+0.1}$ \\
NGC 3089 & \texttt{phabs*pow} & 28.55/27 & $7.1_{-3.5}^{+5.4}\times10^{22}$ & $2.7_{-0.8}^{+1.1}$ & $-11.7_{-0.1}^{+0.2}$ \\
NGC 3227 & \texttt{pow} & 290.09/292 & $<6.1\times10^{20}$ & $1.7_{-0.1}^{+0.1}$ & $-10.10_{-0.03}^{+0.03}$ \\
NGC 4151 & \texttt{zbbody + zxipcf*zcutoffpl} & 258.15/328 & $7.8_{-2.0}^{+2.2}\times10^{22}$ & $1.51_{-0.03}^{+0.03}$ & $-9.60_{-0.08}^{+0.06}$ \\
NGC 4180 & \texttt{MYTorus} & 5.40/2 & $7.3_{-1.5}^{+1.7}\times10^{24}$ & $1.5_{-0.2}^{+0.3}$ & $-11.1$ \\
NGC 4235 & \texttt{phabs*pow} & 150.28/181 & $2.7_{-1.4}^{+1.6}\times10^{21}$ & $1.7_{-0.3}^{+0.3}$ & $-11.13_{-0.06}^{+0.06}$ \\
NGC 4388 & \texttt{phabs*pow + const*pow} & 248.06/265 & $4.6_{-0.9}^{+1.0}\times10^{23}$ & $3.2_{-0.9}^{+1.0}$ & $-9.5_{-0.1}^{+0.2}$ \\
NGC 4593 & \texttt{pow} & 304.40/353 & $<2.7\times10^{20}$ & $1.6_{-0.09}^{+0.09}$ & $-10.45_{-0.03}^{+0.03}$ \\
NGC 5290 & \texttt{phabs*pow} & 114.27/145 & $9.5_{-5.3}^{+6.5}\times10^{21}$ & $1.1_{-0.5}^{+0.5}$ & $-10.83_{-0.05}^{+0.05}$ \\
NGC 5506 & \texttt{zxipcf*pexrav} & 257.79/304 & $1.2_{-0.1}^{+0.6}\times10^{22}$ & $1.7_{-0.2}^{+0.3}$ & $-9.97_{-0.03}^{+0.10}$ \\
NGC 5899 & \texttt{phabs*pow} & 51.39/49 & $10.9_{-3.8}^{+5.1}\times10^{22}$ & $2.0_{-0.2}^{+0.2}$ & $-11.1_{-0.1}^{+0.1}$ \\
NGC 6814 & \texttt{pow} & 324.07/350 & $<5.4\times10^{20}$ & $1.6_{-0.1}^{+0.1}$ & $-10.68_{-0.03}^{+0.03}$ \\
NGC 7314 & \texttt{phabs*pow} & 247.97/293 & $8.6_{-1.3}^{+1.4}\times10^{21}$ & $1.9_{-0.1}^{+0.1}$ & $-10.69_{-0.03}^{+0.03}$ \\
NGC 7378 & \texttt{phabs*pow} & 8.94/9 & $5.8_{-2.1}^{+2.2}\times10^{22}$ & $1.7_{-0.1}^{+0.1}$ & $-12.1_{-0.2}^{+1.1}$ \\
NGC 7465 & \texttt{phabs*pow} & 166.46/215 & $8.2_{-2.6}^{+3.0}\times10^{21}$ & $1.6_{-0.2}^{+0.3}$ & $-10.95_{-0.05}^{+0.05}$ \\
NGC 7479 & \texttt{apec + MYTorus} & 15.52/9 & $5.7_{-0.5}^{+0.5}\times10^{24}$ & $2.5_{-0.1}^{+0.1}$ & $-8.8$
\enddata
\label{tab:XRTdata}
\caption{X-ray spectral fitting results. All (formal) confidence intervals and upper limits are 90\% except for the X-ray fluxes, which are $\pm1\sigma$. X-ray fluxes without a confidence interval are from fits not using contemporaneous Swift XRT data. Note that Swift XRT was in an anomaly state for NGC~3786, so we could not use its data.}
\end{deluxetable*}

\subsection{The Fundamental Plane at High Physical Resolution} \label{subsec: highres}
We have shown that, when probed at the exquisite angular resolution of the VLBA, AGN are under-luminous or altogether undetected, at odds with their properties in lower angular resolution VLA data, but potentially more in line with the scaling relation for Galactic X-ray binaries. We can remove the dependence of black hole mass in this relation for a pure comparison between X-ray and radio luminosities and produce a relationship that may better encapsulate the core behavior of AGNs. Figure~\ref{fig:rvsx} compares L$_{6\,cm}$ and L$_{2-10\,keV}$ in our sample, with detections as red points and radio luminosity upper-limits derived from RMS values for non-detections as green points, revealing VLBA observed AGN largely fall in line with measurements for Galactic black holes (GBHs). While the majority of our sample is undetected, these non-detections are significant and so in order to preserve the statistical power of our sample we fold in the information these non-detections carry by performing regression with censored data using the \textsc{asurv} package \citep{1992ASPC...25..245L}. 

$L_{5~\mathrm{GHz}}$ and $L_{2-10~\mathrm{keV}}$ measurements from Table 1 of \citep{2013MNRAS.432.1138P} are also included as blue points. Initially, there appears to be a discrepancy between how both datasets compare to measurements from GBHs. However, separating radio-quiet and radio-loud AGN according to the X-ray radio loudness parameter ($R_X \equiv L_{6~\mathrm{cm}}/L_{2-10~\mathrm{keV}}$ limit ($R_X=-4.5$) from \citet{2003ApJ...583..145T}, we find the eight radio-loud AGN in the \citet{2013MNRAS.432.1138P} sample are the targets which do not strongly agree with the general trend. Additionally, NGC~1052 from our sample is also radio-loud by this definition, likewise separating itself from the general trend, which concurs with its use as an ICRF target and the documented presence of a relativistic radio jet \citep{2019ApJ...874...43L} often present in radio-loud AGN. We note that while GBHs may themselves exhibit bimodality in $L_R/L_X$ \citep[e.g.,][]{2012MNRAS.423..590G}, these modes do not appear to be reflected in the AGN population. Using the two best-fit relations from \citet{2012MNRAS.423..590G}, the GBHs in this work are consistent with one or the other, but both relations predict radio luminosities that are three to four orders of magnitude below the radio-loud relation at the X-ray luminosities of AGNs. Of the two GBH $L_R/L_X$ ``tracks'' presented in \citet{2012MNRAS.423..590G}, the lower track, corresponding to GBHs that are less radio-luminous, is more consistent with the AGNs studied in this work.

As even completely uncorrelated fluxes can become highly correlated when multiplied by random distances spanning several orders of magnitude, we fit the FP in terms of flux as well as luminosity normalized by BH mass. The results are shown in Figure~\ref{fig:rvsxflux}, which show that the correlation between the 2--10~keV and 6~cm fluxes and BH mass-normalized luminosities is weak, and that the values derived using the high physical resolution VLBA data overlap with the values seen in XRBs, which are systematically lower than AGNs observed in lower-resolution studies.


Finally, we compared the radio luminosities to the observed 14--195~keV luminosities from the 105-month BAT spectra alone, using the fluxes calculated in Section~\ref{subsec: x-ray analysis} and provided in Table~\ref{tab:batonly}. For the X-ray data from the \citet{2003MNRAS.345.1057M} sample, which was mostly taken from published literature using a wide variety of X-ray observatories, we multiply the 2-10~keV luminosities by a factor of 1.6 to estimate the 14-195~keV luminosities, a factor derived from a power-law with a spectral index of 2. The results are shown in the right panel of Figure~\ref{fig:rvsx}. As can be seen, the 2--10~keV fluxes calculated using our simultaneous Swift XRT observations are consistent with those inferred from the BAT data, indicating that apparent under-luminosity seen in the radio between VLA and VLBA data is not in fact due to an over-luminosity at softer X-rays due to contamination from extra-nuclear sources.

\begin{deluxetable}{llrr} \label{tab:batonly}
\tablehead{\colhead{Object} & \colhead{Model} & \colhead{$\chi^2/\mathrm{dof}$} & \colhead{$\log_{10}\left(F_\mathrm{14-195\,keV}\right)$} \\ [-0.2cm]
\colhead{} & \colhead{} & \colhead{} & \colhead{[erg~cm$^{-2}$~s$^{-1}$]}}
\startdata
NGC 1052 & pow & 4.87/5 & $-10.505^{+0.036}_{-0.046}$ \\
NGC 1068 & phabs*pow & 2.51/4 & $-10.515^{+0.031}_{-0.047}$ \\
NGC 1320 & pow & 3.31/5 & $-10.872^{+0.067}_{-0.121}$ \\
NGC 2110 & cutoffpl & 4.16/4 & $-9.496^{+0.004}_{-0.006}$ \\
NGC 2782 & pow & 1.45/5 & $-10.921^{+0.068}_{-0.136}$ \\
IC 2461 & pow & 3.26/5 & $-10.722^{+0.011}_{-0.025}$ \\
NGC 2992 & pow & 1.43/5 & $-10.487^{+0.035}_{-0.046}$ \\
NGC 3079 & phabs*cutoffpl & 2.78/3 & $-10.550^{+0.053}_{-0.008}$ \\
NGC 3081 & phabs*cutoffpl & 1.53/3 & $-10.125^{+0.012}_{-0.026}$ \\
NGC 3089 & pow & 0.76/5 & $-11.156^{+0.072}_{-0.248}$ \\
NGC 3227 & phabs*cutoffpl & 4.34/3 & $-9.990^{+0.015}_{-0.010}$ \\
NGC 3786 & pow & 2.1/5 & $-10.836^{+0.058}_{-0.095}$ \\
NGC 4151 & cutoffpl & 7.11/4 & $-9.249^{+0.003}_{-0.002}$ \\
NGC 4180 & pow & 9.14/5 & $-10.847^{+0.074}_{-0.139}$ \\
NGC 4235 & pow & 4.29/5 & $-10.414^{+0.028}_{-0.034}$ \\
NGC 4388 & pow & 2.18/4 & $-9.569^{+0.005}_{-0.005}$ \\
NGC 4593 & cutoffpl & 0.87/4 & $-10.112^{+0.034}_{-0.013}$ \\
NGC 5290 & pow & 2.54/5 & $-10.835^{+0.054}_{-0.076}$ \\
NGC 5506 & phabs*cutoffpl & 2.93/3 & $-9.700^{+0.013}_{-0.004}$ \\
NGC 5899 & pow & 2.03/5 & $-10.688^{+0.037}_{-0.053}$ \\
NGC 6814 & cutoffpl & 2.18/4 & $-10.258^{+0.050}_{-0.012}$ \\
NGC 7314 & cutoffpl & 0.44/4 & $-10.335^{+0.046}_{-0.018}$ \\
NGC 7378 & pow & 2.21/5 & $-10.863^{+0.079}_{-0.170}$ \\
NGC 7465 & pow & 3.5/5 & $-10.722^{+0.055}_{-0.083}$ \\
NGC 7479 & pow & 5.31/5 & $-10.774^{+0.058}_{-0.096}$
\enddata
\caption{Best-fit \textsc{xspec} models for the Swift BAT data \texttt{from the 105-month catalog \citep{2018ApJS..235....4O}}, with observed 14--195~keV fluxes.}
\end{deluxetable}

\subsection{The Nature of Sub-Parsec-Scale Radio Emission} \label{subsec: nature}
With the results detailed in Section~\ref{subsec: vlavlba} and Section~\ref{subsec: highres}, it is clear that there is a resolution dependence of the perceived radio luminosity of nearby AGNs, that at sub-parsec scales the radio emission is considerably less luminous, and that the FP does not appear to capture the relationship between the black hole mass, X-ray luminosity, and radio emission at these scales. It is therefore worthwhile to explore what physical mechanisms may be behind the radio emission that we do detect on mas scales.

In higher-frequency studies \citep[e.g., 22~GHz][]{2019MNRAS.488.4317B,2020MNRAS.492.4216S}, there may be some mixture of synchrotron and thermal bremsstrahlung (free-free) emission, which begins to dominate for star-forming galaxies above $\sim30$~GHz \citep{1992ARA&A..30..575C}. However, for our observations, which were taken at 5~GHz, it is more likely that the emission is synchrotron-dominated. This is not necessarily true, however, if the radio emission is ``young'', which induces a spectral curvature, or turnover at lower frequencies. One means of distinguishing these emission mechanisms is by calculating the brightness temperature:
\begin{equation} \label{eq: Tb}
T_b = \frac{4 \ln(2) c^2 S_\nu}{2 \pi k_B \theta_A \theta_B},
\end{equation}
\noindent where $S_\nu$ is the specific intensity over the solid angle, and $\theta_A$ and $\theta_B$ are the full widths at half power of the major and minor axes of an elliptical Gaussian beam. 

\noindent We find that the brightness temperatures calculated from our VLBA detections range from \texttt{$10^{6.1}$~K to $10^{9.7}$~K}, indicating a likely dominant synchrotron source. One potential source of compact emission, thermal free-free emission from accretion disk winds, may be significantly present in NGC~1068 and NGC~2992, which have brightness temperatures of $10^{6.1}$~K and $10^{6.3}$~K, respectively. NGC~1068 exhibits extreme line-of-sight absorption $N_\mathrm{H}\sim10^{25}$~cm$^{-2}$ \citep{2017ApJS..233...17R}, which may scatter or absorb direct non-thermal AGN emission \citep{1997Natur.388..852G, Gal04}. For the remaining objects, $T_b\geq10^7$, suggesting either jet-associated or corona-associated synchrotron. While our observations sample down to the parsec or sub-parsec scale, it is possible that sub-parsec jets may exist, as was recently found for a small fraction (4\%) of BAT-selected AGNs using other VLBI observations \citep{2019MNRAS.488.4317B}. However, the AGNs in that study are more distant, with radio luminosities several orders of magnitude larger than the sample studied here. Indeed, \citet{2019MNRAS.488.4317B} find that the $L_R/L_X$ relation of their sample spans a range of $10^{-3.5}-10^{-1.5}$, indicating jet-dominated radio emission. By contrast, our sample has $L_R/L_X$ between $\sim10^{-7}-10^{-4}$, more in line with expectations from coronal emission, which scales as approximately $L_R/L_X\sim10^{-5}$ \citep{2008MNRAS.390..847L, 2019NatAs...3..387P}.

If the radio emission is coronal in origin, then one might expect that the relationship between the X-ray luminosity, radio luminosity, and black hole mass would be more pronounced (i.e., a tighter FP), on account not only of the X-ray and radio emission originating in the same physical structure, but also that structure existing only a few gravitational radii from the black hole \citep[e.g.,][]{2013ApJ...769L...7R}. However, we have found that the correlation between the X-ray and parsec-scale radio emission in hard X-ray selected AGNs is rather poor, and is not improved by the introduction of the black hole mass. Indeed, the majority of our sample is not even detected in the radio on parsec scales, in stark contrast with lower resolution studies \citep[e.g.,][]{2020MNRAS.492.4216S}, with $L_R/L_X$ well below the fiducial $10^{-5}$ scaling for pure coronal emission \citep{2008MNRAS.390..847L, 2019NatAs...3..387P}. We suggest that there may be ``radio silent'' AGNs: AGNs that do not even produce significant coronal synchrotron radio emission. Deeper follow-up studies of these undetected sources will prove critical to confirming this, but in either case our results underscore the need to better understand and characterize the physical properties of the corona.




\section{Conclusions} \label{section: Conclusions}

In this work, we have explored the simultaneous X-ray (2--10~keV) and radio (C-band; 5~GHz) properties of a complete sample of AGNs with hard X-ray (14--195~keV) luminosities above $10^{42}$~erg~s$^{-1}$ from the local volume out to $40$~Mpc, within the declination limits of $-30\arcdeg$ to $+60\arcdeg$. Our X-ray observations, which come from Swift~XRT, have allowed us to fit the X-ray spectra of our targets, factoring in prior information from the long-term average BAT spectra, obtaining absorption-corrected 2--10~keV luminosities for comparison with the radio to X-ray luminosity relation $L_R/L_X$, as well as the Fundamental Plane of black hole activity, the latter of which seeks to unify the radio and X-ray accretion properties of both Galactic and supermassive black holes by including the black hole mass as a third parameter. Our radio observations, which come from the VLBA, were designed to achieve good $uv$ coverage, and have allowed us to probe the radio emission of these AGNs in the sub-parsec regime, at the cost of sensitivity to large-scale extended radio emission. These data have therefore enabled a systematic study of the contemporaneous relationship between the X-ray corona and the origin of core radio emission in AGNs, a relationship that is not currently well understood. Our conclusions are as follows:

\begin{enumerate}
    \item AGN fall out of the Fundamental Plane when radio fluxes are measured on mas (sub-parsec) scales. The difference in flux between VLBA and VLA observations 
    suggests that measurements made with the VLA are contaminated by additional extra-nuclear radio emission that is resolved out in VLBA observations, and that this extended extra-nuclear radio emission is responsible for the Fundamental Plane. 
    
    \item Radio-quiet AGN share a similar relationship between $L_\mathrm{5~GHz}$ and $L_\mathrm{2-10~keV}$ with nearby Galactic black holes. Radio-loud AGN ($R_X <-$4.5) do 
    not share this relation and do not appear to have Galactic black hole analogs.
    
    \item These results are consistent with a fundamental bimodality in $L_R/L_X$ for AGNs, supporting their classification as either radio-loud or radio-quiet. Caution is therefore warranted when extrapolating physical models of nearby AGNs to their higher redshift, radio-bright counterparts. 
    
    \item Despite the depth of our VLBA observations ($\mathrm{RMS}\sim20$~$\mu$Jy), the majority of our sample is undetected, with hard upper limits ($3\sigma$) on the $L_R/L_X$ relation as low as $<10^{-7}$, below even expectations from their coronal X-ray emission, which scales as approximately $L_R/L_X\sim10^{-5}$, suggesting AGNs that are effectively ``radio-silent''. These findings underscore the need for further dedicated study of the physics of the X-ray corona.

\end{enumerate}

\acknowledgments
We thank Claudio Ricci for helpful suggestions regarding the X-ray analysis.

The National Radio Astronomy Observatory is a facility of the National Science Foundation operated under cooperative agreement by Associated Universities, Inc. The authors acknowledge use of the Very Long Baseline Array under the US Naval Observatory's time allocation.   This work supports USNO's ongoing research into the celestial reference frame and geodesy. This work made use of data supplied by the UK Swift Science Data Centre at the University of Leicester. The images used throughout this work were made using APLpy. \citep{2012ascl.soft08017R}.

\vspace{5mm}
\facilities{VLBA, Swift, VLA, EVLA}

\software{\textsc{aips}, Astropy~\citep{2013A&A...558A..33A}, \textsc{casa}, \textsc{xspec}
          }

\appendix

\section{Notes on Individual Objects} \label{section: Notes on Individual Objects}
\subsection*{NGC 2782} \label{subsection: NGC 2782}
\begin{figure}
\centering
\includegraphics[width=0.5\columnwidth]{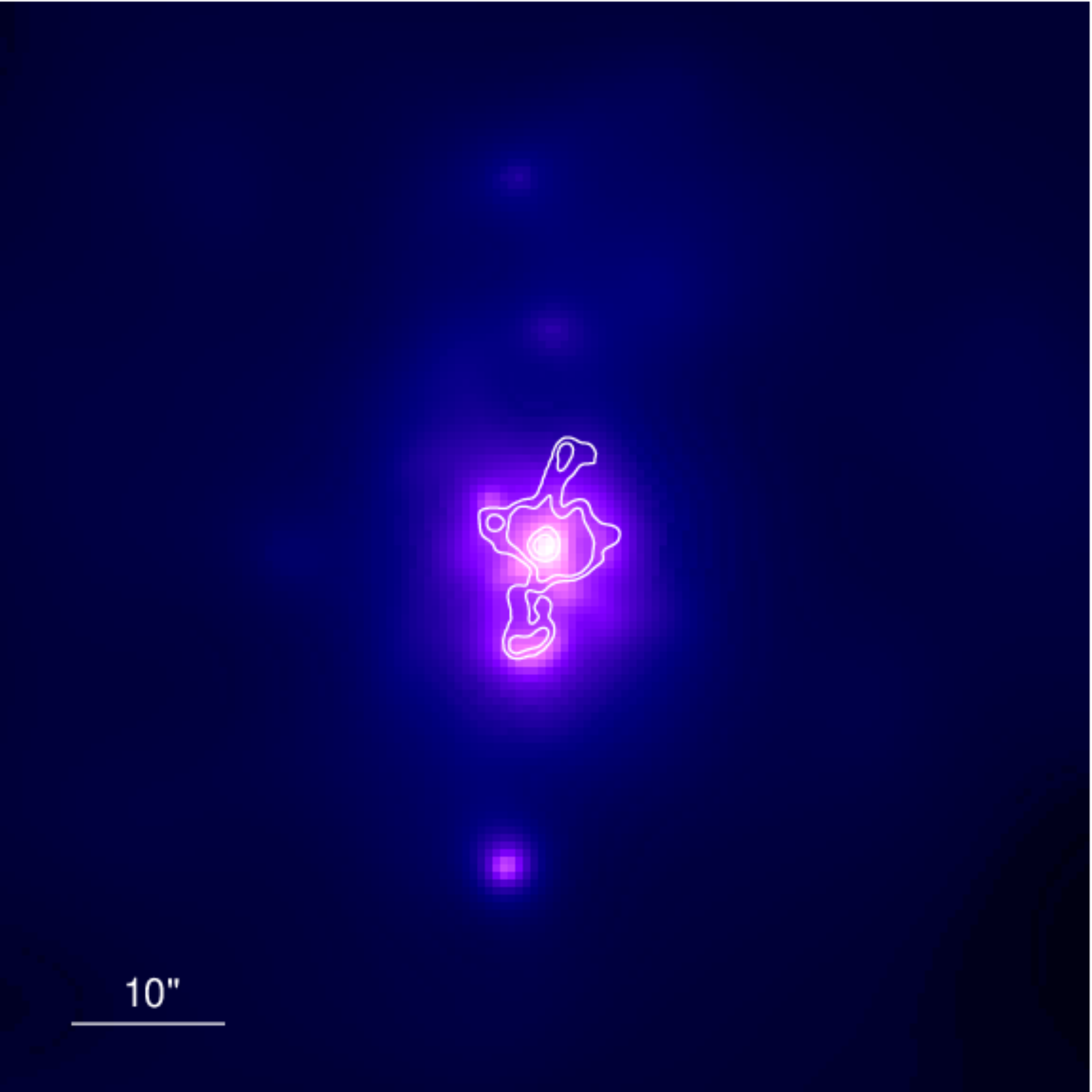}
\caption{NGC 2782 CXO 0.5--10~keV image (adaptively smoothed). For reference we also include C-band (6~GHz) contours from an archival B/A-configuration VLA observation, showing how the extended X-ray emission coincides with the extended radio continuum. Image is approximately the same size as the XRT extraction aperture.}
\label{fig: ngc2782cxo}
\end{figure}

This object was suggested to be Compton-thick by \citet{2006A&A...450..933Z}, who fit Chandra X-ray Observatory (CXO) data with a plasma emission component plus a high-energy reflection model. \citet{2007A&A...468..129T}, using the same data, notice an Fe~K$\alpha$ line (6.4~keV) at high significance with an equivalent width (EW) of $\sim1.5$~keV. We use an APEC model to account for the plasma component, setting the temperature and abundance equal to the values in \citet{2006A&A...450..933Z}, and fit the spectrum using MYTorus. We obtain a column density $N_\mathrm{H}$ value of $1.3^{+4.7}_{-1.3}\times10^{23}$~cm$^{-2}$, suggesting that the source is heavily absorbed if not Compton-thick. However, the relative normalization between the scattered (``reflection'') continuum and the intrinsic power-law component is over four orders of magnitude different between time-averaged BAT spectrum and the XRT data, despite their intrinsic power-law fits varying in normalization only marginally, which is unphysical. Conversely, if we assume that the relative normalization of the scattered continuum is constant between epochs, the best-fit $N_\mathrm{H}$ is unchanged but the intrinsic power-law normalization corresponding to the XRT data is $\sim25\%$ that of the BAT data, indicating considerable variability. Setting the strength of the scattered continuum to be fixed between the BAT and XRT epochs, which allows for the possibility of a time delay between a change in the intrinsic power-law continuum and a change in the scattered continuum, leaves this result unchanged, likely due to the weakness of the scattered continuum at low $N_\mathrm{H}$ values. However, the complex nature of the soft X-ray spectrum allows for a wide range of possible $N_\mathrm{H}$ values, as we have seen. To mitigate this issue, we downloaded and reprocessed the archival CXO data discussed above (Obs~ID 3014; 2002 May 17; PI: Stevens) using \textsc{ciao}, version 4.10 with CALDB~4.7.8. Using a $r=1\farcs5$ aperture aperture centered on the AGN, and defining an $1\farcs4<r<3\farcs5$ annular background region immediately surrounding the AGN to sample the diffuse soft X-ray emission that contaminates the low angular resolution XRT data (Figure~\ref{fig: ngc2782cxo}), we extracted the isolated AGN spectrum using \texttt{specextract}. With the AGN isolated, we were able to delete the diffuse plasma APEC component, fitting only the MYTorus model. We find a hydrogen column density of $N_\mathrm{H} = 1.4^{+0.3}_{-0.1}\times10^{23}$~cm$^{-2}$, consistent with the best-fit value using the XRT data, but with significantly tighter confidence bounds. We therefore classify NGC~2782 as not being Compton-thick, contrary to the findings of \citet{2006A&A...450..933Z}. With $N_\mathrm{H}$ frozen to this value, we re-fit the XRT$+$BAT data, finding that the power-law normalization at the XRT epoch is $26\%$ that of the BAT data, in agreement with our initial finding of variability between the epochs. Finally, we note that, in addition to diffuse soft X-ray emission, there are 5 unresolved X-ray sources within the XRT aperture, consistent with being X-ray binaries (XRBs) or background AGNs (Figure~\ref{fig: ngc2782cxo}). We extracted the background-isolated summed spectrum for all 5, finding it to be well-fit ($\chi^2/\mathrm{dof}=1.03$) with a power-law component with $\Gamma=1.7$ and $N_\mathrm{H}=6.5\times10^{20}$~cm$^{-2}$. The total 2--10~keV flux is $3.5\times10^{-14}$~erg~cm$^{-2}$~s$^{-1}$, $\sim3\%$ of the flux from the absorbed power-law component fit to the XRT data. We therefore do not consider these sources to be a significant contaminant in our results.

\subsection*{NGC 3079} \label{subsection: NGC 3079}
\begin{figure}
\centering
\includegraphics[width=0.5\columnwidth]{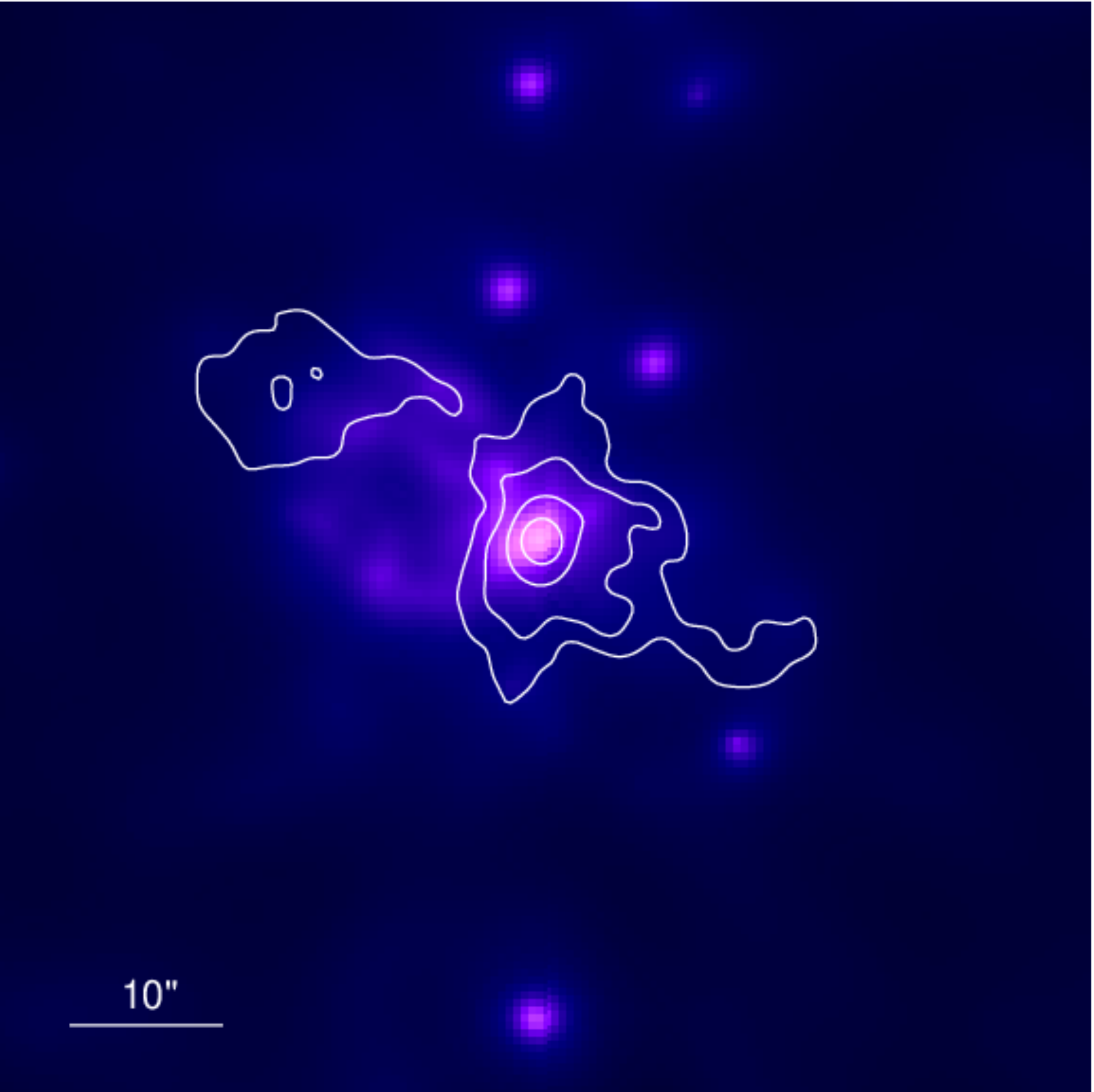}
\caption{NGC 3079 CXO 0.5--10~keV image (adaptively smoothed). For reference we also include L-band (1.4~GHz) contours from an archival A-configuration VLA observation, showing extended radio emission ahead of the loop of soft X-ray emission.}
\label{fig: ngc3079cxo}
\end{figure}

\begin{figure}
\centering
\includegraphics[angle=-90,width=0.5\columnwidth]{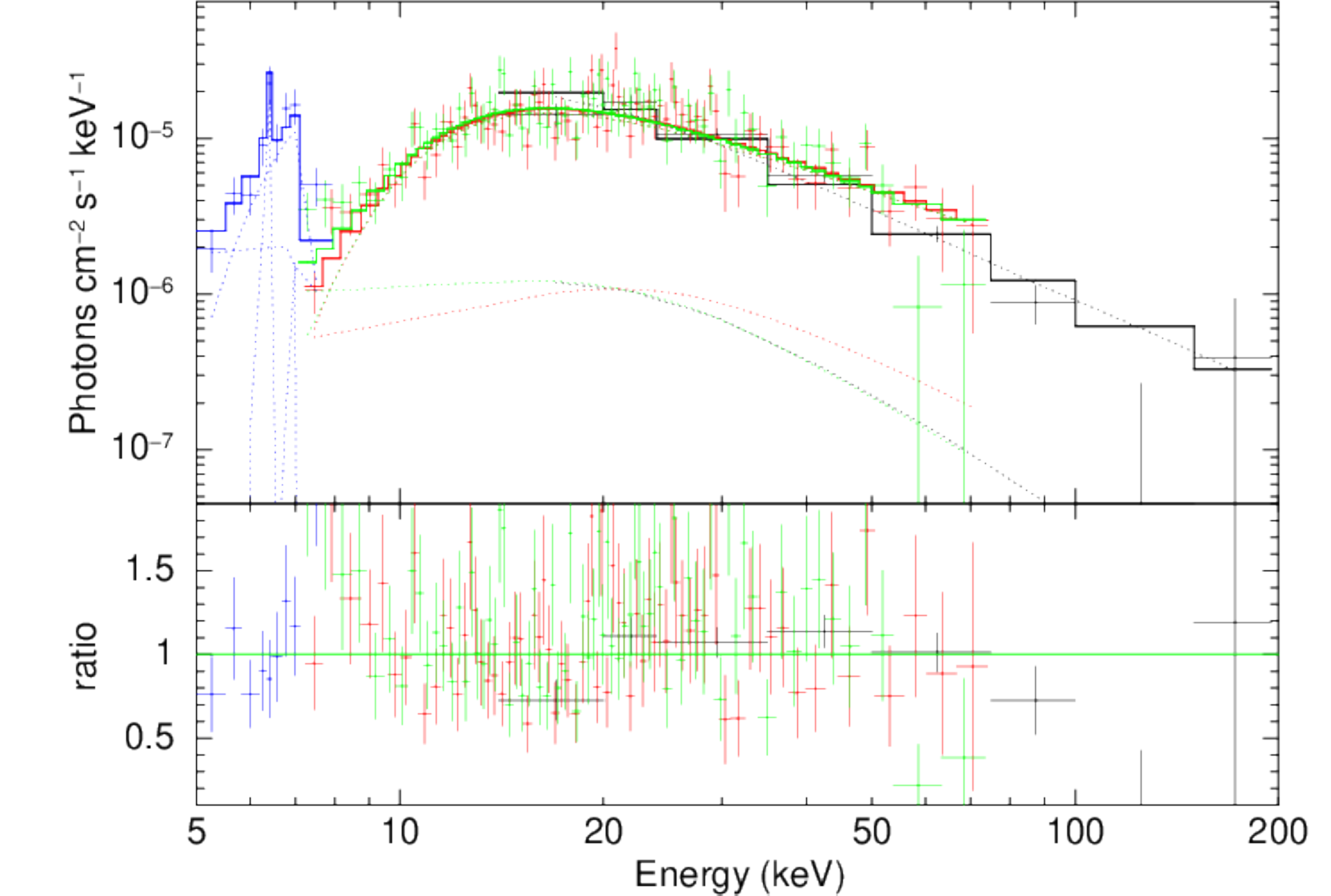}
\caption{NGC 3079 X-ray spectrum CXO data in blue, NuSTAR FMPA/FPMB data in red and green, and BAT data in black. The angular resolution of CXO has allowed us to confine the AGN emission at soft X-ray energies and fit the Fe~K$\alpha$ line, and the sensitivity of NuSTAR has allowed us to obtain a more precise estimate of $N_\mathrm{H}$.}
\label{fig: ngc3079_cxo_nsr_bat}
\end{figure}

This is a known Compton-thick object \citep[e.g.,][]{2017ApJS..233...17R}, and has an extensive nuclear superbubble \citep[e.g.,][]{2019ApJ...873...27L} that contributes significantly to the X-ray spectrum at soft ($<2$~keV) energies. As with NGC~2782, we isolate the nuclear X-ray emission using CXO data (Obs~ID 19307; 2018 January 30; PI: Li), using the immediate vicinity of the AGN for the background spectrum. We also found an archival NuSTAR observation (Obs~ID 60061097002; 2013 November 12; PI: Harrison) taken as part of the NuSTAR extragalactic surveys \citep{2016ApJ...831..185H}. Using MYTorus, we found a good fit ($\chi^2/\mathrm{dof} = 197.95/155$) with $N_\mathrm{H}=8.5^{+1.2}_{-1.0}\times10^{24}$~cm$^{-2}$, although we have to fit the CXO and NuSTAR data respectively above 4 and 7~keV to avoid residual contamination from the diffuse X-ray emission seen in the CXO image (Figure~\ref{fig: ngc3079cxo}). We show this spectrum in Figure~\ref{fig: ngc3079_cxo_nsr_bat}. We note that we find evidence of variability in the intrinsic power-law spectral index $\Gamma$ between the NuSTAR and BAT data, with the former having $\Gamma=1.7^{+0.2}_{-0.2}$ and the latter having $\Gamma=2.2^{+0.1}_{-0.1}$.

\begin{figure}
\centering
\includegraphics[angle=-90,width=0.5\columnwidth]{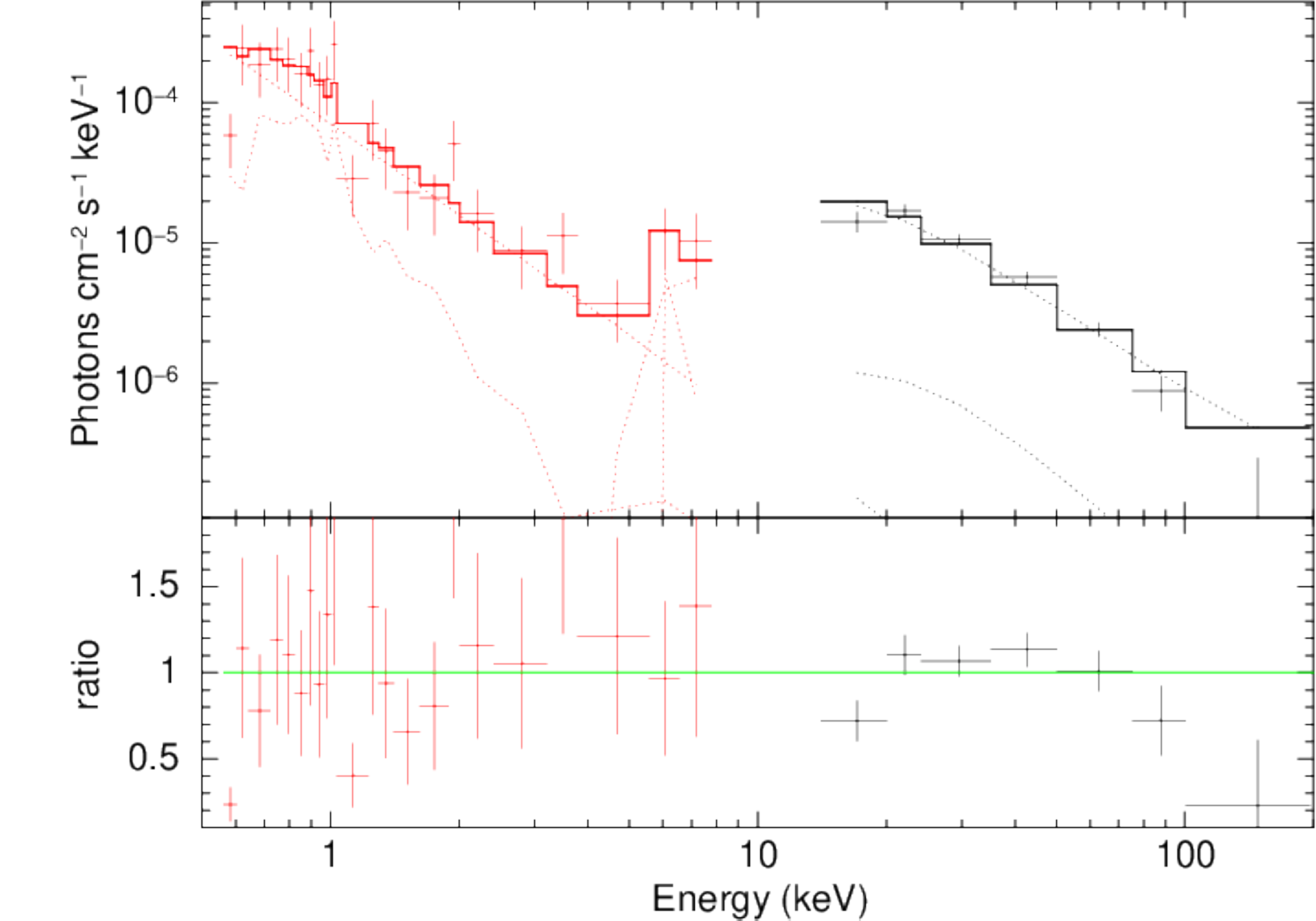}
\caption{NGC 3079 X-ray spectrum with contemporaneous XRT data in red and BAT 105-month data in black.}
\label{fig: ngc3079_xrt_bat}
\end{figure}

We then fit the XRT and BAT data separately, holding the MYTorus component parameters fixed to the best-fit values found above, and adding before the MYTorus component an Astrophysical Plasma Emission Code \citep[APEC;][]{2001ApJ...556L..91S}\footnote{\url{http://www.atomdb.org}} model, using the default \texttt{angr} abundances \citep{1989GeCoA..53..197A}. With the exception of the normalization, we keep all APEC parameters frozen to those found in \citet{2019ApJ...873...27L} by using the \texttt{vapec} variant. Although the intrinsic power-law continuum is of course not detectable in the XRT data for this bone fide Compton-thick object, precluding a measurement of the unabsorbed 2--10~keV flux contemporaneous with our VLBA observations, the XRT data do reveal the presence of additional X-ray emission above $\sim1$~keV not attributable to the plasma emission component. We find that this additional emission is consistent with an optically-thin scattered continuum component, which can be interpreted as intrinsic power-law emission originating along an unobscured direction being scattered off free electrons back into our line of sight. We model this component as a power-law with parameters held fixed to the intrinsic power-law component fit to the BAT data, with a coefficient to account for the scattering fraction. We find this value to be $\sim0.1\%$. We show this spectrum in Figure~\ref{fig: ngc3079_xrt_bat}.

\subsection*{NGC 3081} \label{subsection: NGC 3081}
\begin{figure}
\centering
\includegraphics[width=0.5\columnwidth]{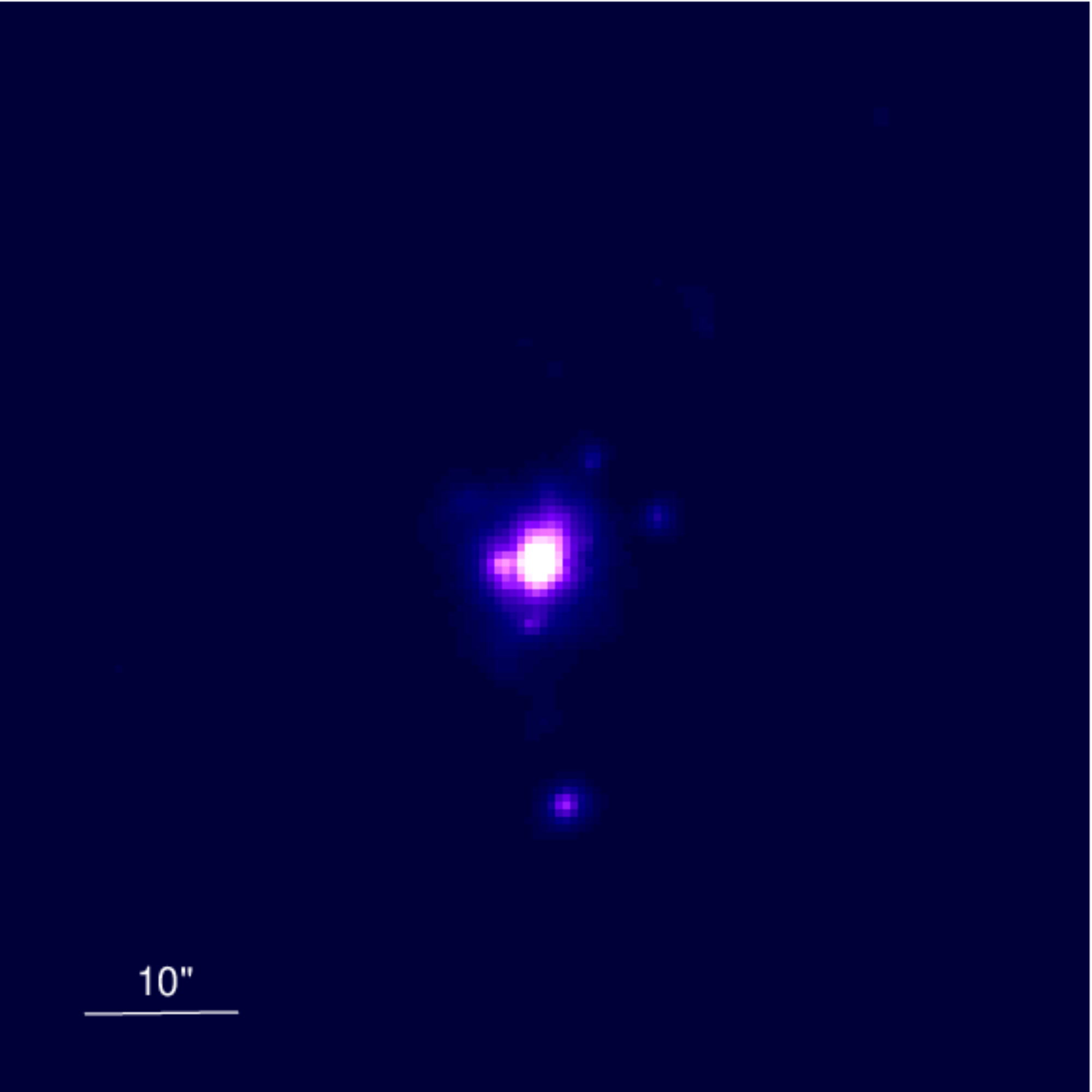}
\caption{NGC 3081 CXO 0.5--10~keV image (adaptively smoothed). The VLA C-band contours lie entirely within the bright central source and so have been omitted.}
\label{fig: ngc3081cxo}
\end{figure}

\begin{figure}
\centering
\includegraphics[angle=-90,width=0.5\columnwidth]{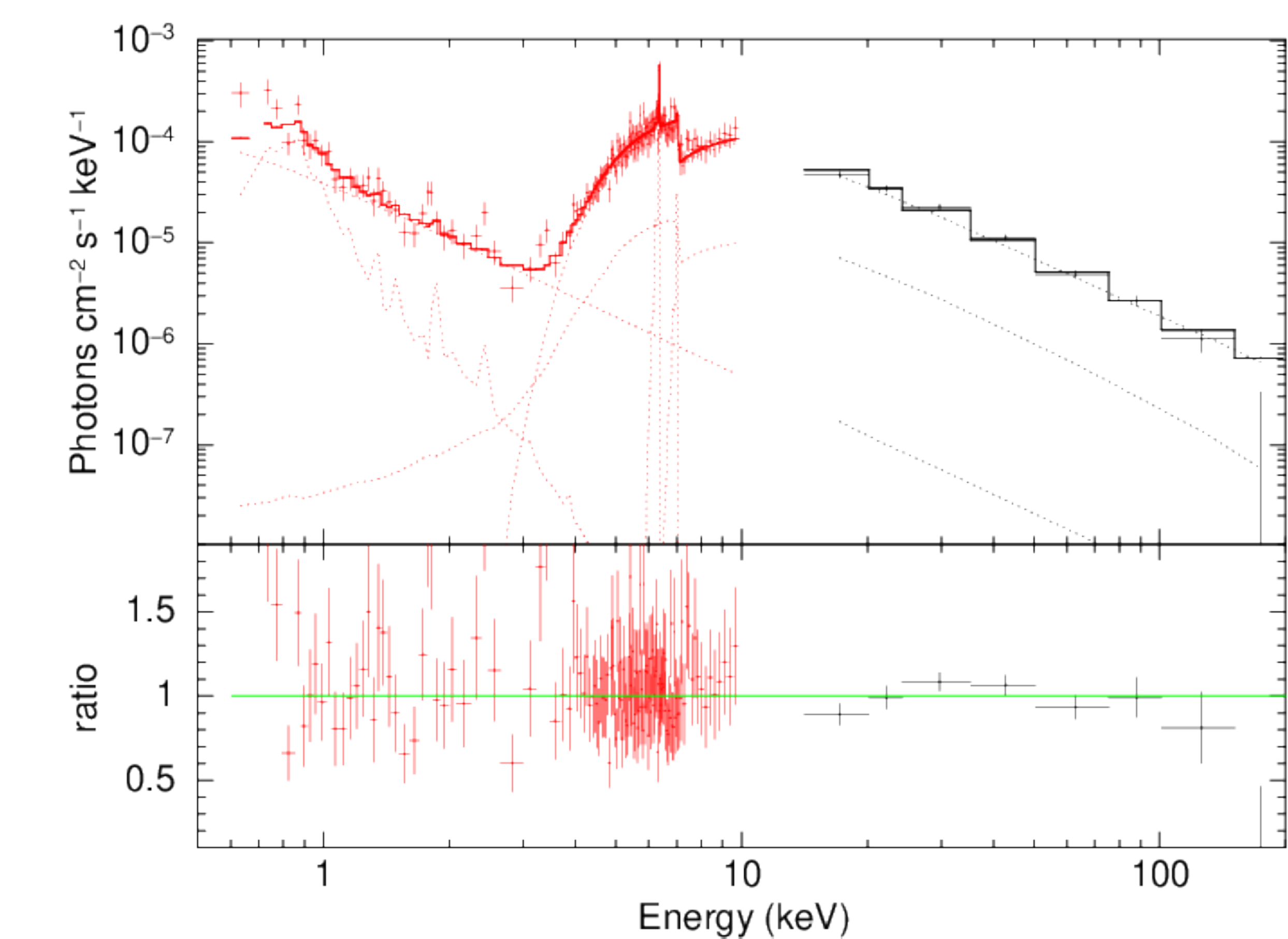}
\caption{NGC 3081 X-ray spectrum with CXO data in red and BAT 105-month data in black.}
\label{fig: ngc3081_cxo_bat}
\end{figure}

One archival CXO observation exists for this object (Obs~ID 20622; 2018 January 24; PI: Maksym). The high angular resolution X-ray imaging enabled by these data reveal a compact source with little diffuse emission and few nearby XRBs (Figure~\ref{fig: ngc3081cxo}). A joint analysis of the CXO and BAT spectrum gives an excellent fit ($\chi^2/\mathrm{dof} = 141.93/141$) with an APEC thermal plasma model appended to MYTorus, consistent with the findings of \citet{2011ApJ...729...31E} who performed a joint fit with Suzaku and Swift BAT 22-month catalog data \citep{2010ApJS..186..378T}. However, by allowing the spectral index corresponding to the CXO data to vary we find that this fit produces an unusually hard spectral index of $\Gamma_\mathrm{CXO}=0.9^{+0.5}_{-0.4}$, which given the uncertainties may be due to the relative insensitivity of the CXO data to the intrinsic power-law continuum in the presence of heavy absorption. Tying $\Gamma$ to the best-fit value of the BAT data group $\Gamma=1.96$ is disfavored by an $F$-test, which gives $p=1.5\times10^{-3}$. Setting $\Gamma_\mathrm{CXO}$ to the $90\%$ upper confidence limit of $1.4$ gives $\chi^2/\mathrm{dof}=144.71/142$ and $N_\mathrm{H}=6.1^{+0.4}_{-0.3}\times10^{23}$~cm$^{-2}$, within 0.2 and 0.1 dex of the values respectively reported in \citet{2011ApJ...729...31E} and \citet{2017ApJS..233...17R}. We show the best-fit spectrum in Figure~\ref{fig: ngc3081_cxo_bat}

\subsection*{NGC 4151} \label{subsection: NGC 4151}
The X-ray spectrum of NGC 4151 is known to be complex, containing a rich array of emission lines at soft ($<1$~keV) energies \citep[e.g.,][]{2004MNRAS.350....1S}, as well as significant spectral variation \citep[e.g.,][]{2019ApJ...884...26Z}. The BAT spectrum exhibits downward curvature at the highest energies suggestive of reflection, and we initially fit the BAT spectrum with the exponentially cut off power-law reflection model \texttt{pexrav} \citep{1995MNRAS.273..837M}. Holding the best-fit parameters fixed, we added the XRT data and appended the partial covering, partially ionized absorber component \texttt{zxipcf} as was used for NGC~4151 in \citet{2016MNRAS.458.2454L}. To account for unresolved line emission below $<1$~keV, we appended a simple black body component, but we do not attribute to its best-fit parameters any physical significance. We found that the fit is insensitive to the reflection scaling factor, with the best-fit value being statistically consistent with zero. We therefore replaced the \texttt{pexrav} component with a power-law with a high energy cutoff (\texttt{zcutoffpl}). The best-fit parameters yield $N_\mathrm{H}=1.9^{+0.4}_{-1.3}\times10^{23}$~cm$^{-2}$ and $\Gamma=1.54^{+0.04}_{-0.04}$.

\subsection*{NGC 4180} \label{subsection: NGC 4180}
\begin{figure}
\centering
\includegraphics[angle=-90,width=0.5\columnwidth]{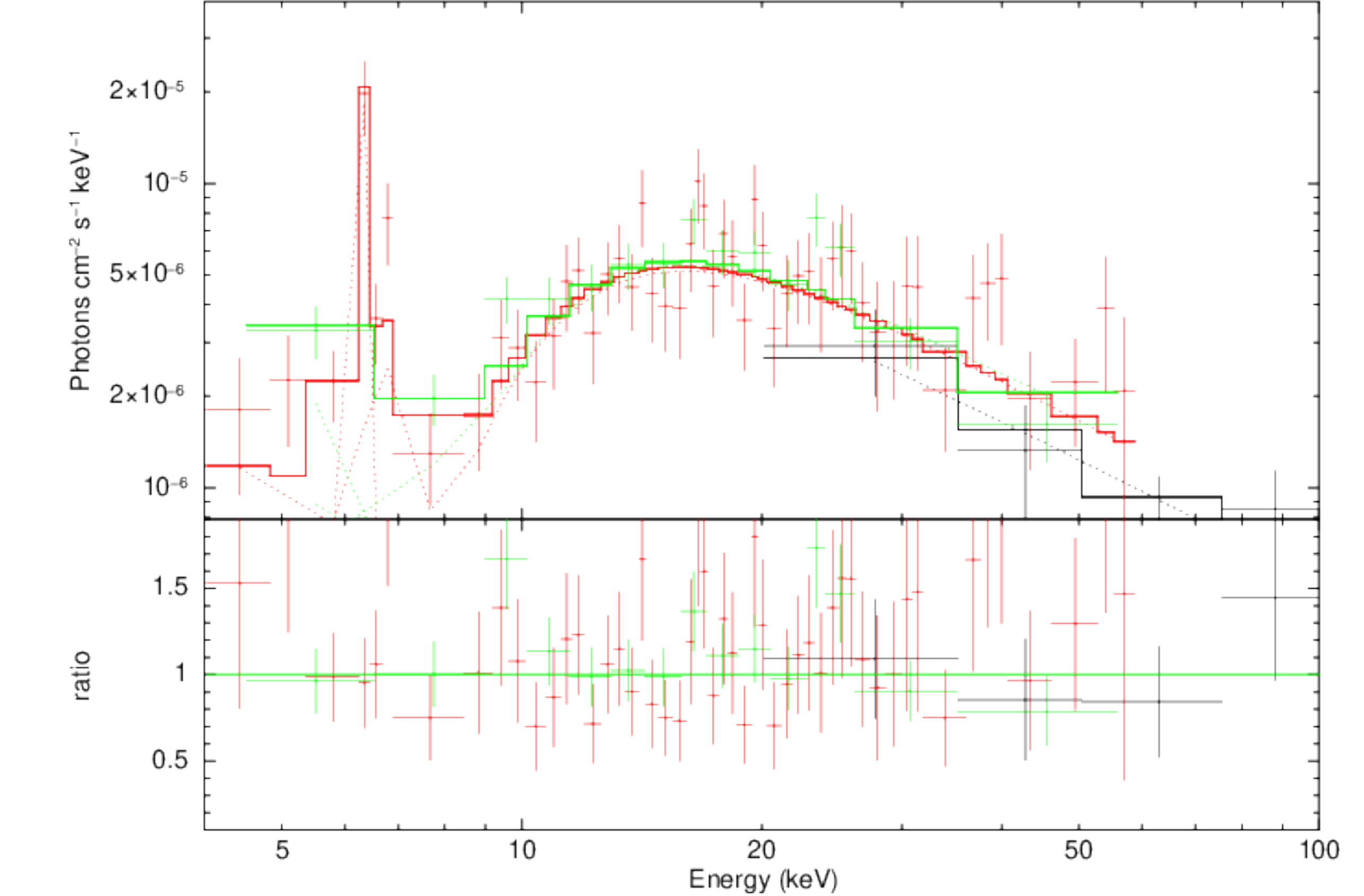}
\caption{NGC 4180 X-ray spectrum with NuSTAR FPMA/FPMB (red/green) and BAT (black) data.}
\label{fig: ngc4180_nsr_bat}
\end{figure}

One CXO observation exists for this object (Obs~ID 9438; 2008 November 16; PI: Walter), but the object is exceptionally faint, with only 8 counts within a $1\farcs5$ aperture centered on the AGN. Using an archival NuSTAR observation (Obs~ID 60201038002), we performed a joint fit with the BAT 105-month spectrum using MYTorus. We find an excellent fit ($\chi^2/\mathrm{dof}=107.31/107$ with $N_\mathrm{H}=7.3^{+1.7}_{-1.5}\times10^{24}$~cm$^{-2}$, indicating that NGC~4180 is a bona fide Compton-thick AGN. We note that this result is somewhat in tension with \citet{2017ApJS..233...17R}, who find $N_\mathrm{H}=1.4^{+0.6}_{-0.6}\times10^{24}$~cm$^{-2}$ using a more phenomenological model and BAT 70-month catalog data \citep{2013ApJS..207...19B}. We show the best-fit spectrum in Figure~\ref{fig: ngc4180_nsr_bat}.

\subsection*{NGC 5506} \label{subsection: NGC 5506}
As with NGC 4151, a downward spectral curvature at high energies is present in the X-ray spectrum of NGC~5506, suggesting reflection or a high energy cutoff, and a partially ionized X-ray absorber is more consistent with the data than a simple photoelectric absorption model (i.e., \texttt{phabs}). However, unlike NGC~4151 we find that the BAT spectrum is significantly better fit with a \texttt{pexrav} reflection model than a \texttt{zcutoffpl} cutoff power-law model ($\chi^2_\nu = 0.44$ versus 3.5), and the latter yields an unusually hard spectral index of $\Gamma=0.8$ while the former yields $\Gamma=1.8$.

\subsection*{NGC 7378} \label{subsection: NGC 7378}
The contemporaneous XRT data for this object, ObsID~00088600003, is exceptionally faint, with only 5 spectral data counts. Consequently, the XRT and BAT data do not preclude the possibility that this source is Compton-thick, and a naive fit using the MYTorus model gives $N_\mathrm{H}\sim3\times10^{24}$~cm$^{-2}$, with a 90\% confidence lower limit of $2\times10^{23}$~cm$^{-2}$ and the upper limit pegged at $10^{25}$~cm$^{-2}$. To resolve this ambiguity, we used archival NuSTAR \citep{2013ApJ...770..103H} FPMA, FPMB data of NGC~7378, ObsID 60464202002 (2018 December 22; 20.3~ks; PI: Harrison), taken as part of the NuSTAR extragalactic surveys \citep{2016ApJ...831..185H}, for which we extracted data products using the interactive analysis feature on the Space Science Data Center website.\footnote{\url{http://nustar.ssdc.asi.it}}

\begin{figure}
\centering
\includegraphics[angle=-90,width=0.5\columnwidth]{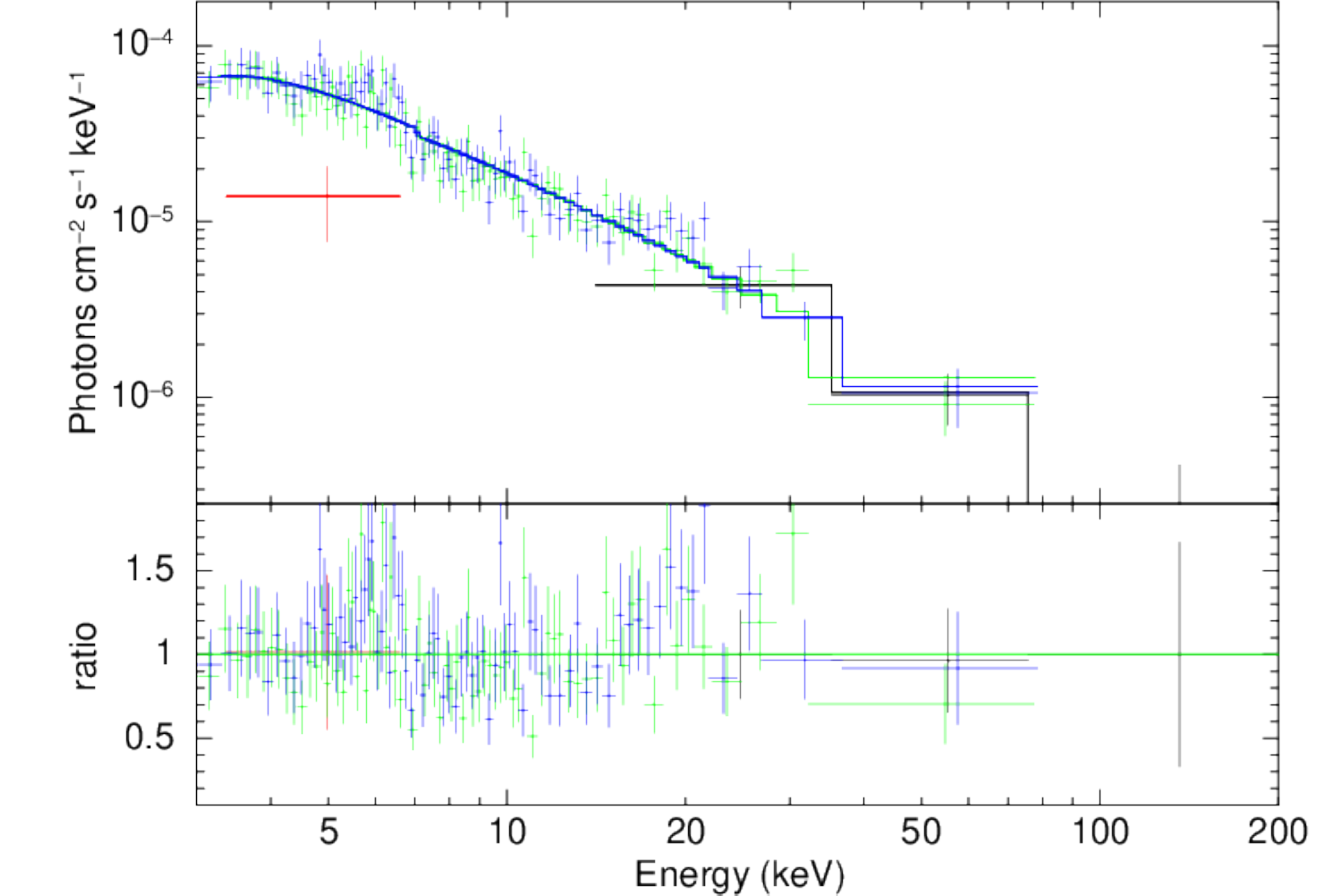}
\caption{NGC~7378 X-ray spectrum with contemporaneous XRT data in red, BAT 105-month data in black, and NuSTAR FPMA/FPMB in green and blue, respectively. The XRT data are significantly fainter than the BAT/NuSTAR data, indicating potential variability at soft X-rays.}
\label{fig: ngc7378}
\end{figure}

After including the NuSTAR data, we find that the X-ray spectrum of NGC~7378 is well-described by an absorbed power-law with $N_\mathrm{H}=5.8^{+2.2}_{-2.1}\times10^{22}$~cm$^{-2}$ and $\Gamma=1.7^{+0.1}_{-0.1}$. The model normalizations of the power-law fits to the BAT and NuSTAR data are consistent with a $\sim10\%$ instrumental calibration uncertainty; however the power-law normalization of the XRT data is considerably lower, being $\sim26\%$ of the value of the NuSTAR data normalization (Figure~\ref{fig: ngc7378}), with a 90\% upper limit of $54\%$. This indicates potential variability between the NuSTAR and XRT epochs, which are separated by 4.5~months, and so we calculate the 2--10~keV flux corresponding to the XRT data normalization while holding the other spectral parameters to the values determined from the fit to all four datasets.

\subsection*{NGC 7479} \label{subsection: NGC 7479}
\begin{figure}
\centering
\includegraphics[width=0.5\columnwidth]{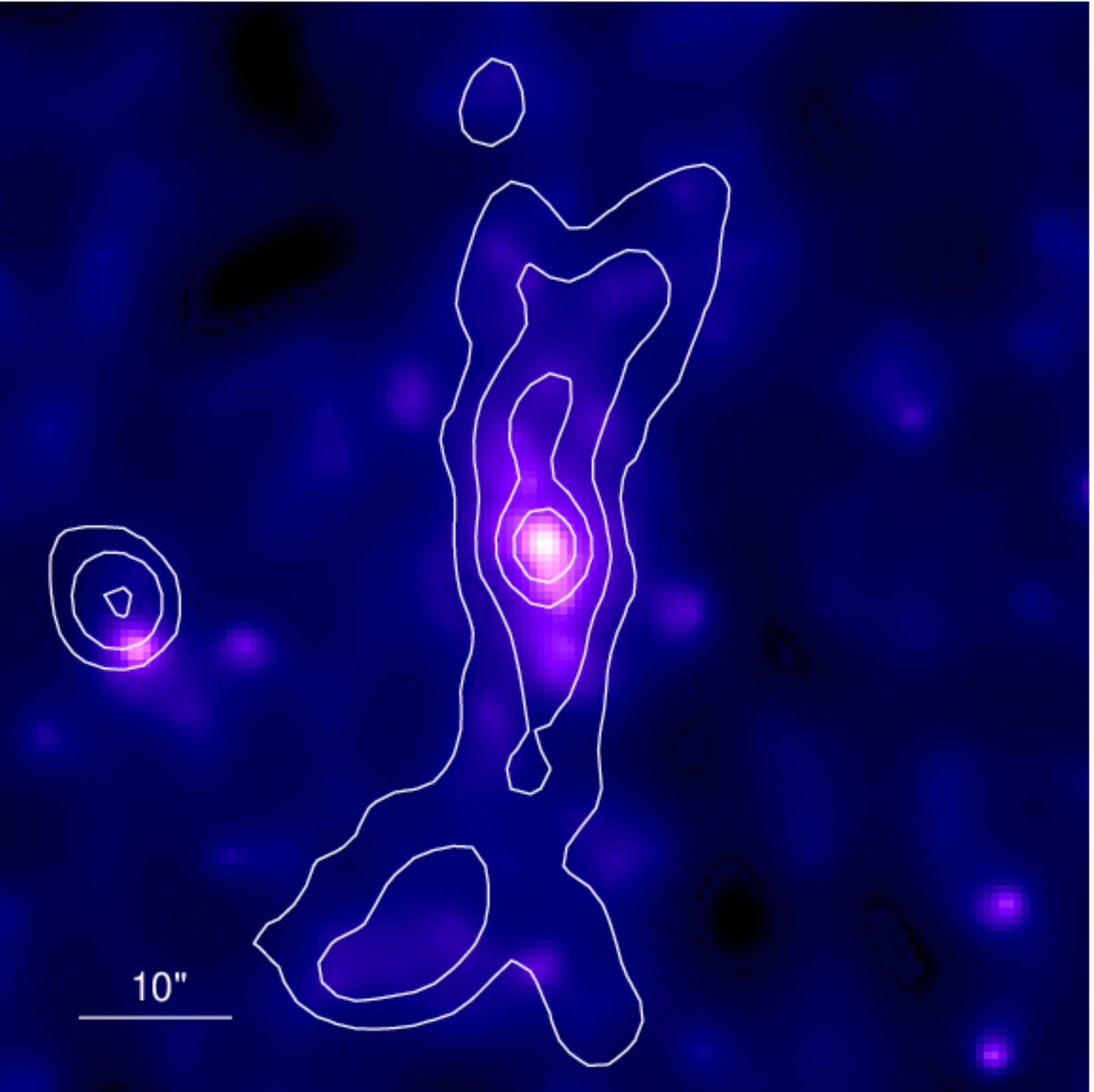}
\caption{NGC 7479 CXO 0.5--10~keV image (adaptively smoothed). For reference we also include C-band (6~GHz) contours from an archival C/C-configuration VLA observation, showing the extended radio morphology coinciding with diffuse soft X-rays emission.}
\label{fig:ngc7479cxo}
\end{figure}

\begin{figure}
\centering
\includegraphics[angle=-90,width=0.5\columnwidth]{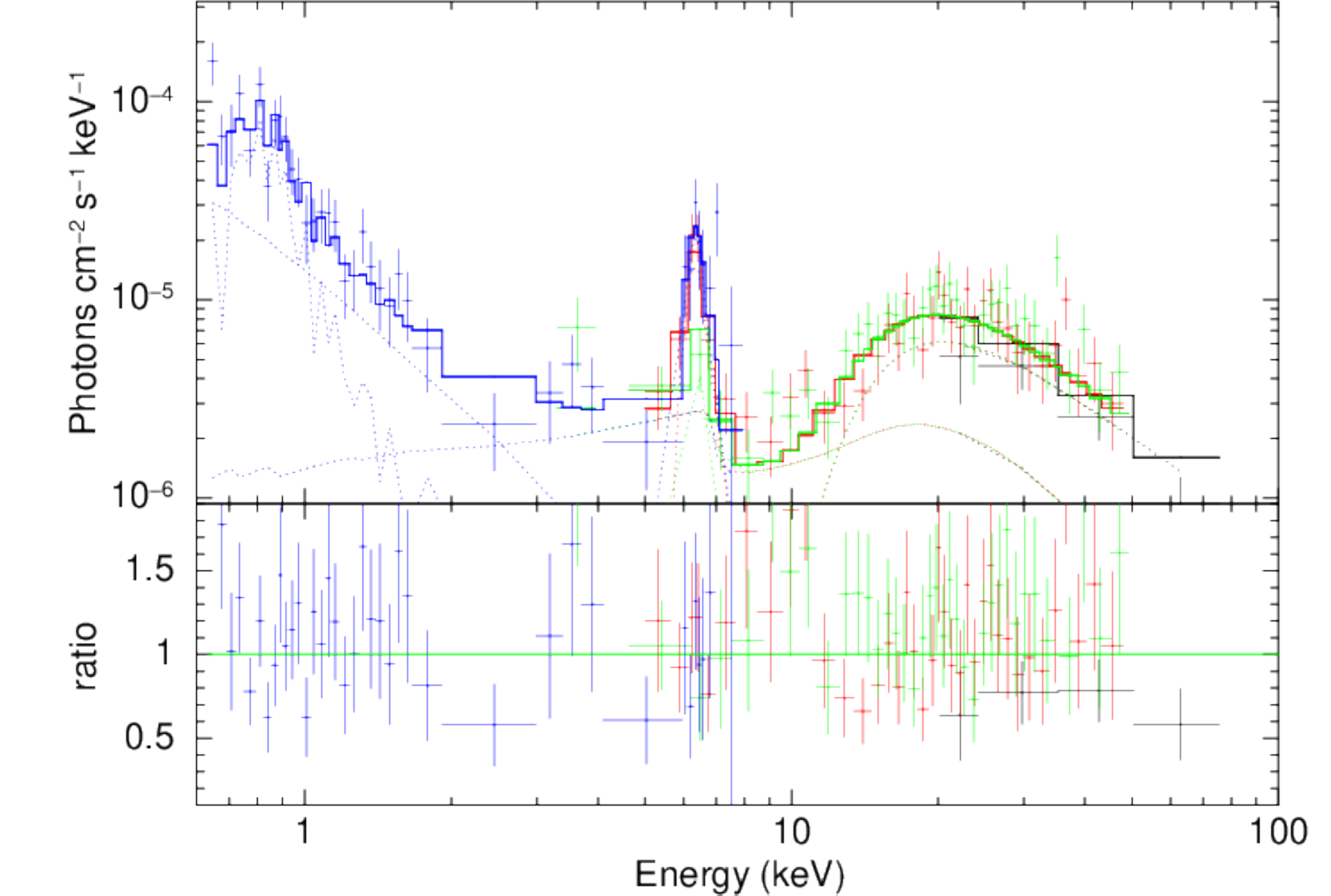}
\caption{NGC 7479 X-ray spectrum.}
\label{fig:ngc7479cxo_nsr_bat}
\end{figure}

This object is listed as borderline Compton-thick ($N_\mathrm{H}=1.4\times10^{24}$~cm$^{-2}$) by \citet{2017ApJS..233...17R}. We reprocessed an archival NuSTAR observation (Obs~ID 60201037002; 2016 May 12; PI: Ricci) of NGC~7479 and jointly fit its BAT 105-month data using MYTorus above 3~keV, finding an excellent fit ($\chi^2/\mathrm{dof}=76.99/77$) with $\Gamma=2.47^{+0.12}_{-0.11}$ and $N_\mathrm{H}=5.7^{+0.5}_{-0.5}\times10^{24}$~cm$^{-2}$, for a torus inclination angle of $\theta=79^{\circ}$, where $0^{\circ}$ is face-on. NGC~7479 is therefore fully Compton-thick, and we do not expect our XRT observations to have any sensitivity to the intrinsic power-law continuum. However, we briefly explore the soft X-ray properties of this object by downloading and reprocessing an archival CXO observation (Obs~ID 11230; 2009 August 11; PI: Pooley). The high resolution CXO observation reveals extensive diffuse emission near the AGN, as shown in Figure~\ref{fig:ngc7479cxo}. We extract the soft X-ray spectrum using an aperture matched to that of NuSTAR. We find that the joint CXO, NuSTAR, and BAT spectrum is well-fit ($\chi^2/\mathrm{dof}=151.04/134$ with the addition of a single APEC plasma component with $kT=0.7$~keV. The intrinsic power-law spectral index and column density are not significantly affected by the addition of the CXO data, although curiously the relative normalization of the Fe~K$\alpha$ line in the CXO data is required to be $\sim4$ times higher than in the other data. We show the fit to the full X-ray spectrum in Figure~\ref{fig:ngc7479cxo_nsr_bat}.


\bibliography{framex}

\end{document}